\documentclass[12pt, letterpaper]{article}

\usepackage{graphicx}
\usepackage[letterpaper,ignoreall,top=2.65cm,bottom=2.7cm,left=2.65cm,right=2.65cm,foot=1cm,head=1.5cm]{geometry}
\usepackage[OT1]{fontenc}
\usepackage[super,sort&compress]{natbib}

\begin{document}

\title{A Little Bit of Carbon Can do a Lot for Superconductivity in H$_3$S}


\author{
Xiaoyu Wang, Tiange Bi, Katerina P. Hilleke \\ 
\small Department of Chemistry, State University of New York at Buffalo, Buffalo, NY 14260-3000, USA \\ [1ex]

Anmol Lamichhane \\ 
\small Department of Physics, University of Illinois Chicago, Chicago, IL 60607-7059, USA \\ [1ex]

Russell J. Hemley$^\ast$ \\ 
\small Department of Physics, University of Illinois Chicago, Chicago, IL 60607-7059, USA \\
\small Department of Chemistry, University of Illinois Chicago, Chicago, IL 60607-7059, USA \\ [1ex]

Eva Zurek\thanks{\raggedright Correspondence should be addressed to: rhemley@uic.edu; ezurek@buffalo.edu}
 \\ 
\small Department of Chemistry, State University of New York at Buffalo, Buffalo, NY 14260-3000, USA \\ [1ex]

}

\maketitle

\begin{abstract}
Recently, room temperature superconductivity was measured in a carbonaceous sulfur hydride material whose identity remains unknown. Herein, first-principles calculations are performed to provide a chemical basis for structural candidates derived by doping H$_3$S with low levels of carbon. Pressure stabilizes unusual bonding configurations about the carbon atoms, which can be six-fold coordinated as CH$_6$ entities within the cubic H$_3$S framework, or four-fold coordinated as methane intercalated into the H-S lattice, with or without an additional hydrogen in the framework. The doping breaks degenerate bands, lowering the density of states at the Fermi level ($N\textsubscript{F}$), and localizing electrons in C-H bonds. Low levels of CH$_4$ doping do not increase $N\textsubscript{F}$ to values as high as those calculated for $Im\bar{3}m$-H$_3$S, but they yield a larger logarithmic average phonon frequency, and an electron-phonon coupling parameter comparable to that of  $R3m$-H$_3$S.  The implications of carbon doping on the superconducting properties are discussed.
\end{abstract}

\newpage

\section*{Introduction}
The decades old quest for a room-temperature superconductor has recently come to fruition. Inspired by Ashcroft's predictions that hydrogen-rich compounds metallized under pressure could be phonon-mediated high-temperature superconductors \cite{Ashcroft:2004a,Zurek:2009c}, synergy between experiment and theory has led to remarkable progress \cite{Zurek:2018m,Zurek:2018d,Livas:2021a}. A superconducting critical temperature, $T\textsubscript{c}$, of 203~K near 150~GPa was reported for H$_3$S \cite{Drozdov:2015a}, followed by a $T\textsubscript{c}$ of 260~K near 200~GPa in LaH$_{10}$ \cite{Somayazulu:2019-La,Drozdov:2019-La}. Recently, a carbonaceous sulfur hydride superconductor with a measured $T\textsubscript{c}$ of 288~K at 267~GPa was discovered \cite{Snider:2020a}, but many questions remain unanswered about this material. Perhaps the most pressing of these questions is: ``What is  the composition and structure of the phase, or phases, responsible for  the remarkably high $T\textsubscript{c}$?'' Recent x-ray diffraction (XRD) studies suggest the material is derived from the Al$_2$Cu structure type up to 180~GPa \cite{Bykova:2021,Lamichhane:2021}, but its evolution upon further compression where the $T\textsubscript{c}$ is highest has  not yet been determined experimentally. The reported $T\textsubscript{c}$ verses pressure \cite{Snider:2020a}  shows evidence for a transition near 200 GPa, although the scatter in the data are consistent with a continuous change in $T\textsubscript{c}$. On the other hand, the XRD results suggest  that this change in $T\textsubscript{c}$ could arise from the collapse of the observed lower symmetry orthorhombic to higher symmetry superconducting phases.\cite{Lamichhane:2021}

To understand the nature of carbonaceous sulfur hydride, it is useful to review the work leading to the discovery of the initial high-$T\textsubscript{c}$ H$_3$S superconductor \cite{Yao-S-review:2018}. Synthesis of novel (H$_2$S)$_2$H$_2$ van der Waals compounds at pressures up to 40 GPa \cite{Strobel:2011a} inspired the computational search for additional H-S phases that might be stable and potentially superconducting at megabar pressures.\cite{Li:2014,Duan:2014} An $Im\bar{3}m$ symmetry H$_3$S phase, which can be described as a body 
centered cubic sulfur lattice with H atoms lying midway between adjacent S atoms ($T\textsubscript{c}$=191-204~K at 200~GPa), was predicted to be stable above 180~GPa. An analogous lower-symmetry $R3m$ phase with asymmetric H-S bonds was preferred at pressures down to 110~GPa ($T\textsubscript{c}$=155-166~K at 130~GPa) \cite{Duan:2014}. Experiments on  the H-S system confirmed the maximum $T\textsubscript{c}$ (203~K) near the expected pressures (155~GPa),  leading to the proposal that the synthesized structure was the predicted $Im\bar{3}m$ phase \cite{Drozdov:2015a}. Subsequently XRD measurements largely confirmed the predicted cubic structure \cite{Einaga:2016,Minkov:2020a}. However, the synthesis conditions were found to dictate the product that formed, and $T\textsubscript{c}$s as low as 33 K were measured. Thus, reproducing the synthesis of the cubic phase proved difficult \cite{Guigue:2017,Goncharov:2016a}, and more recently altogether new structures have been reported \cite{Laniel:2020a}. 

These observations suggest that a number of H$_x$S$_y$ superconductors can be made. Indeed, 
additional peaks observed in XRD measurements have been assigned to possible secondary phases 
\cite{Einaga:2016,Goncharov:2016a}. Various stoichiometries including H$_2$S \cite{Li:2014,Akashi:2015-S}, HS$_2$ \cite{Errea:2015a}, H$_4$S$_3$ \cite{Li:2016-S}, and H$_5$S$_2$ \cite{Ishikawa:2016} have been proposed for materials with lower $T\textsubscript{c}$s. Exotic Magn\'{e}li phases with H$_x$S$_{1-x}$ ($2/3<x<3/4$) compositions characterized by alternating H$_2$S and H$_3$S regions with a long modulation whose ratio can be varied to tune the $T\textsubscript{c}$ have also been proposed \cite{Akashi:2016a}. First-principles calculations suggested that H$_2$S self-ionizes under pressure forming a (SH$^-$)(H$_3$S)$^+$ perovskite-type structure \cite{Gordon:2016} that may undergo further deformations to a complex modulated phase 
\cite{Majumdar:S-2017,Majumdar:S-2019}. A $Z=24$ $R\bar{3}m$ symmetry phase whose 
density of states (DOS) at the Fermi level ($E\textsubscript{F}$) was predicted to be lower than that of $R3m$ 
and $Im\bar{3}m$ H$_3$S was computed to be more stable than these two phases between 110-165~GPa 
\cite{Verma:2018a}. The role of quantum nuclear and anharmonic effects on the $R3m 
\rightarrow Im\bar{3}m$ transition and $T_c$ has been investigated \cite{Errea:2016,Bianco:2018a}, as has the response of the Fermi surface to uniaxial strain \cite{Liu-strain}. 

Turning to the C-S-H ternary, initial crystal structure prediction calculations conducted 
prior to the discovery of the carbonaceous sulfur hydride superconductor considered stoichiometric C$_x$S$_y$H$_z$ compositions with relatively high carbon dopings  \cite{Zurek:2020b,Sun:2020a}. These studies identified metastable CSH$_7$ structures that were based on the intercalation of methane into an H$_3$S framework, with maximum $T\textsubscript{c}$s estimated to be 194~K at 150~GPa \cite{Zurek:2020b} and 181~K at 100~GPa \cite{Sun:2020a}. However, the measured structural parameters, $P$-$V$ equations of state \cite{Lamichhane:2021}, and the variation of $T_c$ versus pressure \cite{Snider:2020a} do not match those calculated for these hydride perovskite-like materials.  First-principles calculations employing the virtual crystal approximation (VCA) predicted that remarkably low-level hole-doping resulting from the incorporation of carbon in the parent H$_3$S phase (i.e.\  C$_{0.038}$S$_{0.962}$H$_3$) could increase the $T_c$ up to 288~K \cite{Ge:2020a,Hu:2020}. It was argued that doping tunes the position of $E\textsubscript{F}$, moving it closer to the maximum in the DOS that arises from the presence of two van Hove singularities (vHs). Because vHs increase the number of states that can participate in the electron-phonon-coupling (EPC) mechanism, this effect is known in general to enhance the total coupling strength, $\lambda$, and in turn the $T\textsubscript{c}$ in conventional superconductors. The role that the vHs play in increasing the $T\textsubscript{c}$ in H$_3$S has been studied in detail \cite{quan2016van,sano2016effect,Ortenzi:2016,Akashi:2020}.

Despite the striking success of the VCA model in reproducing theoretically the very high $T\textsubscript{c}$ 
measured for the C-S-H superconductor \cite{Ge:2020a}, this approach does not take 
into account the effect of the doping on the local structure and electronic properties, and its limitations have been discussed \cite{Wang:2021,Guan:2021}. To overcome these limitations we systematically study the role of doping on the thermodynamic and dynamic stability, electronic structure, and geometric properties of phases with doping levels as low as 1.85\%. Three types of substitutions are considered involving S replaced by C, together with different numbers of hydrogens, yielding either six-fold or four-fold coordinate carbon atoms. We find 
that CH$_6$ and CH$_4$ form stable configurations within the dense solid in phases that are dynamically stable at the pressures studied experimentally. Morevoer, doping \emph{decreases} the DOS at $E\textsubscript{F}$ because it breaks degeneracies  and localizes electrons in C-H bonds. Our results illustrate that the rigid band model does not reliably predict the superconducting properties of the doped phases. Finally,  the descriptors associated with superconductivity, such as the DOS at $E_F$ and the logarithmic average phonon frequency are used to identify the C-S-H phase likely to possess the highest $T_c$.

\section*{Results and Discussion}

\subsection*{Octahedrally Coordinated Carbon in C$_x$S$_{1-x}$H$_3$ Phases}

To investigate how different levels of doping affect the kinetic and thermodynamic stability, 
electronic structure, and superconducting properties of H$_3$S, we constructed supercells of the $Im\bar{3}m$ structure where one of the sulfur atoms was replaced by carbon. Calculations were carried out at 270~GPa with 
C$_x$S$_{1-x}$H$_3$ stoichiometries and two different types of coordination environments 
around the dopant atom were considered. In the first the  carbon atom was octahedrally 
coordinated by six hydrogen atoms, and in the second two of these C-H bonds broke resulting 
in a quasi-tetrahedral CH$_4$ molecule. A detailed analysis was carried out on the 
C$_{0.0625}$S$_{0.9375}$H$_{3}$ (CS$_{15}$H$_{48}$) stoichiometry at 270~GPa because 
both geometries were dynamically stable at this pressure (Figure~S2d, S10c), and because this unit 
cell size pushes the limits of the phonon and EPC calculations. 
\begin{figure*}
\begin{center}
\includegraphics[width=0.6\textwidth]{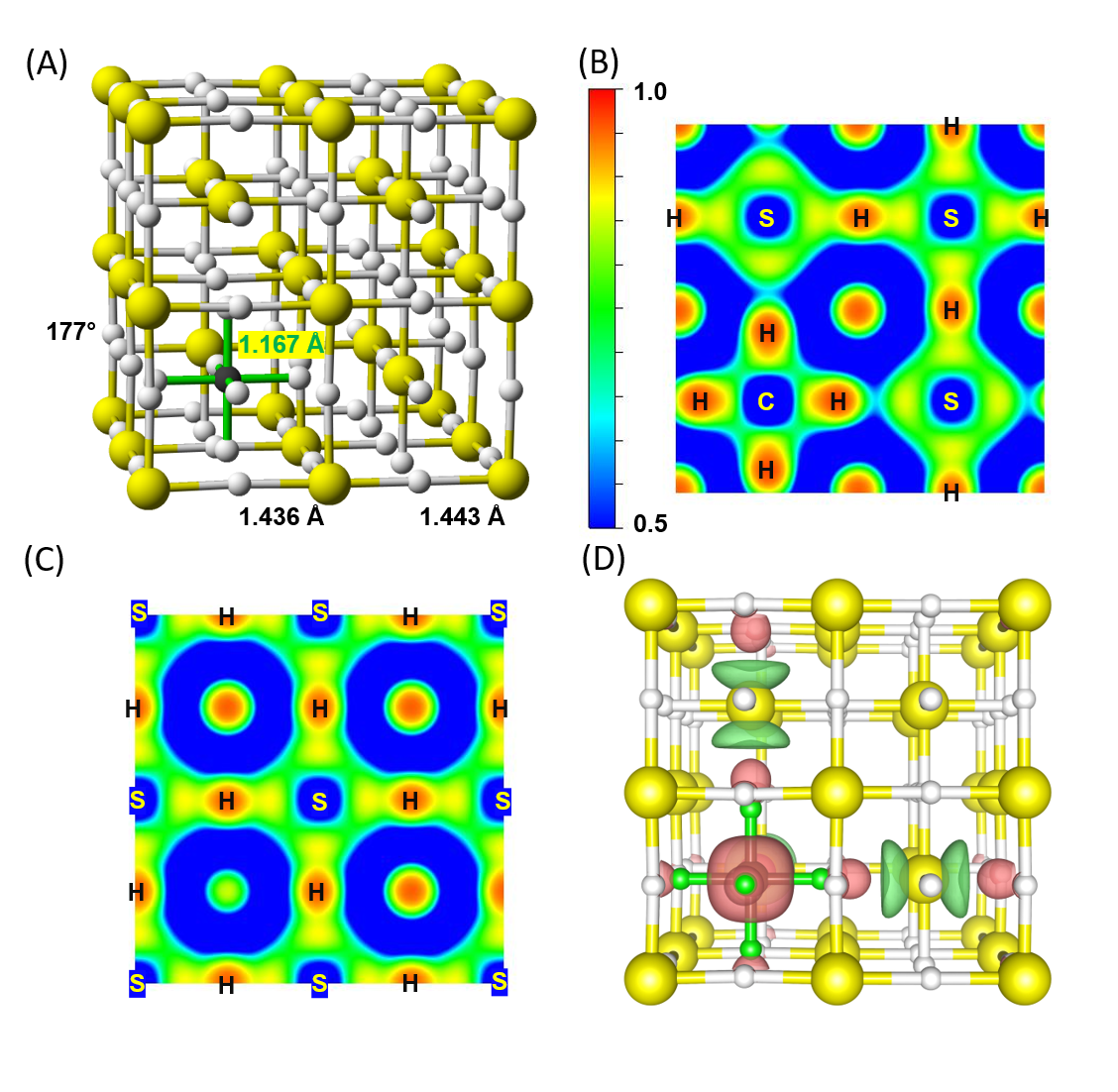}
\end{center}
\caption{(a) Optimized geometry and (b,c) electron localization function (ELF) contour plot of 
C$_{0.0625}$S$_{0.9375}$H$_{3}$ ($O_h$-CS$_{15}$H$_{48}$, space group $Pm\bar{3}m$),  where the 
carbon atom is octahedrally coordinated by hydrogen atoms at 270~GPa. The plane of the 
contour passes through (b) CH$_6$ and (c) H$_3$S, and the range of the isovalues is from 0.5 
(blue/cold) to 1.0 (red/warm). The atoms through which the plane passes are labelled. (d) The 
following electron density difference:  $\rho$(CH$_6$) + $\rho$(S$_{15}$H$_{48}$) - 
$\rho$(CS$_{15}$H$_{48}$) where red denotes a loss of charge and green a gain of charge 
(isovalue=0.005). Sulfur/carbon/hydrogen atoms are yellow/black/white and select geometric 
parameters are provided.
\label{fig:Oh}}
\end{figure*}

The calculated C-H distance in the octahedrally coordinated phase, which we refer to as 
$O_h$-CS$_{15}$H$_{48}$ (Figure\ \ref{fig:Oh}a), measures 1.17~\AA{}. The negative of 
the crystal orbital Hamilton population integrated to the Fermi level (-iCOHP), which can be 
used to quantify the bond strength, is calculated to be 5.11~eV for this bond. The distance 
between the hydrogen atom bonded to carbon and its nearest neighbor sulfur (1.71~\AA{}), is 
significantly larger than the H-S distance found in $Im\bar{3}m$ H$_3$S at this pressure 
(1.45~\AA{}). The weakening of the CH-S bond upon doping is evident in the -iCOHP, which 
decreases from 3.74~eV/bond (in H$_3$S) to 2.00~eV/bond. Despite the non-negligible -iCOHPs between 
the hydrogen atoms bonded to carbon and the nearest neighbor sulfur atoms, the ELF (Figure\ 
\ref{fig:Oh}b,c) does not show any evidence of covalent CH-S bond formation. 
The integrated crystal orbital bond index (iCOBI), which 
is a quantification of the extent of covalent bond formation \cite{Muller:2021}, is calculated to be 0.34 for each S-H 
bond in H$_3$S, indicating a bond order of roughly $1/3$. 
Substituting C in an octahedral coordination environment leaves the S-H iCOBI essentially 
unchanged for bonds distant from the C substitution site, but for CH-S bonds the iCOBI drops 
to 0.21, showing the weakening of this bond. Practically no 
antibonding states are filled in the S-H interaction with H coordinating C (Figure~S6), 
but the increased S-H distance leads to an overall decrease in the magnitude of the iCOBI. The 
C-H bonds in the octahedral CH$_6$ motif possess the largest iCOBI in the system, 0.51. 
Subtracting the charge density of the CS$_{15}$H$_{48}$ structure from a sum of the density 
arising from the neutral CH$_6$ molecule and the neutral H-S framework (Figure\ \ref{fig:Oh}d) 
illustrates that charge is transferred from the CH$_6$ unit into the nearest neighbor sulfur lone 
pairs, which can also be seen in the ELF plot (Figure\ \ref{fig:Oh}b). A Bader analysis, which 
typically underestimates the formal charge, yields a +0.25 charge on CH$_6$, indicating that 
charge redistribution is associated with stabilization of this configuration in the dense structure.

An analysis using the reversed approximation Molecular Orbital (raMO) method, which uses linear
combinations of the occupied crystal orbitals of the system to reproduce
target orbitals, revealed substantial bonding interactions within the CH$_6$
cluster. The reproduced $s$ orbital on the hydrogen atom in this motif contained electron density with $p$-orbital symmetry on the neighboring C atom, indicative of $sp$ bonding.
Similarly, the reproduced $s$ orbital on C strongly interacted with the 
surrounding H atoms, which is evident in its anisotropy as comparison to the more
isotropic sulfur $s$ orbital reproductions (Figure S7).

Theoretical considerations have been key in designing ways to stabilize four-coordinate carbon in novel bonding configurations such as planar tetracoordinate carbon \cite{hoffmann1971theoretical,hoffmann1970planar}.
While a wide variety of molecular compounds containing coordination numbers surpassing 
four, such as carbocations, carboranes, organometallics, and carbon clusters, are also known 
\cite{Olah:1997a,Olah:1987}, octahedrally coordinated carbon is quite unusual. Examples 
include elemental carbon, which has been predicted to become six-fold coordination at 
terapascal pressures \cite{Fahy:1987,Pickard:carbon}, and high pressure Si-C compounds such 
as rock-salt SiC \cite{Daviau:2017a,Hatch:2005a}, and two predicted Si$_3$C phases 
\cite{Gao:Si3C}. More relevant to the C-S-H system are carbon atoms bonded to more than 
four hydrogen atoms such as the nonclassical carbocation, $C_s$-CH$_5^+$, which contains three 
short and two long C-H bonds. It can be viewed as a proton inserted into one of the $\sigma$ C-H bonds within methane, forming a three-center two-electron (3c-2e) bond between one 
carbon and two hydrogen atoms \cite{Marx:1995a}. \emph{Ab initio} calculations for the 
isolated molecule have shown that the minimum energy configuration for the di-carbocation, 
CH$_6^{2+}$, possesses $C_{2v}$ symmetry with two long 3c-2e and two short classic 2c-2e 
bonds, rather than the $O_h$ symmetry CH$_6^{2+}$ geometry 
\cite{Lammertsma:1982,Lammertsma:1983}. Following one of the triply degenerate imaginary 
normal modes, which can be described as a wagging motion along the three sets of H-C-H 
180$^\circ$ angles in the octahedron, leads to the $C_{2v}$ minimum. Turning now to 
hypercoordinated carbon atom in the $O_h$-CS$_{15}$H$_{48}$ model structure, explicit calculation 
of the phonons at the $\Gamma$ point reveals that in the solid state the frequency of 
this same triply degenerate mode is real (calculated frequency of 1712~cm$^{-1}$), and the 
vibration is coupled with the motions of the hydrogens in the H$_3$S lattice. Thus, the 
stabilization of the octahedral molecular complex is facilitated by weak interactions with the 
host lattice. Notably, the calculated C-H bond length in CH$_6^{2+}$ obtained at the HF/6-311+G(2d,p) level of theory,  is nearly identical to that of $O_h$-CS$_{15}$H$_{48}$ at 
270~GPa (both $\sim$1.17~\AA{}). That the carbon weakly interacts with the host lattice is supported by calculations where the carbon (or sulfur) is replaced with neon. The optimized $O_h$-NeS$_{15}$H$_{48}$ structure is dynamically stable at these pressures, with a calculated Ne-H distance of 1.37~\AA{} (Figure S30).

\subsection*{Quasi-Tetrahedrally Coordinated Carbon in C$_x$S$_{1-x}$H$_3$ Phases}
\begin{figure*}[h!]
\begin{center}
\includegraphics[width=0.6\textwidth]{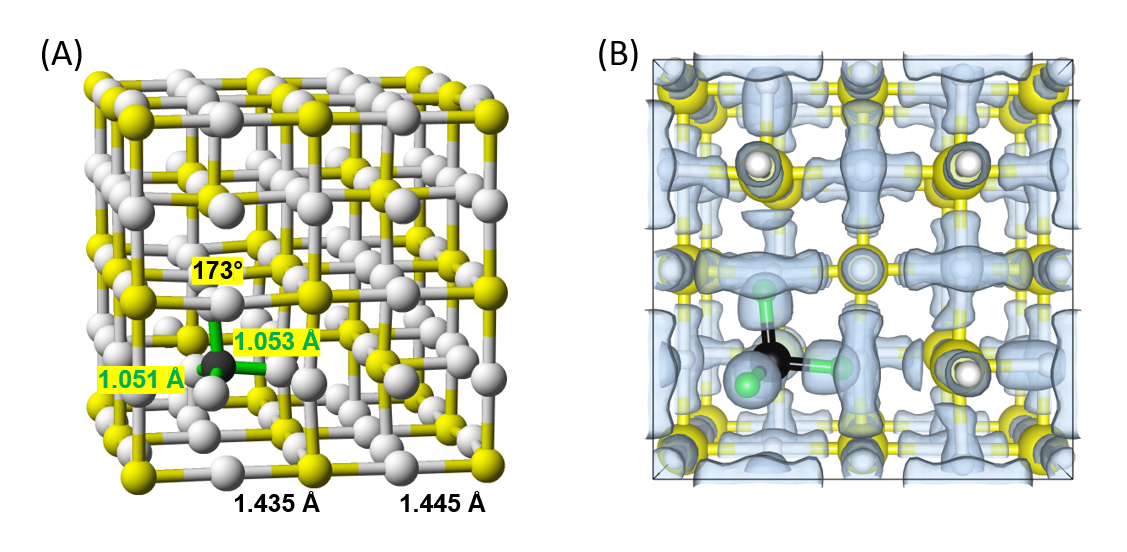}
\end{center}
\caption{(a) Optimized geometry and (b) electron localization function (ELF; isovalue=0.8) of 
the CH$_4$-based C$_{0.0625}$S$_{0.9375}$H$_{3}$ ($T_d$-CS$_{15}$H$_{48}$, space group 
$Amm2$) in which the carbon atom is quasi-tetrahedrally coordinated by hydrogen atoms at 
270~GPa. Sulfur/carbon/hydrogen atoms are yellow/black/white and select geometric 
parameters are provided.
\label{fig:Td}}
\end{figure*}

When placed in a cube a tetrahedral methane molecule can retain its symmetry only if it its hydrogens point toward four corners of the  cube. In the phases studied here such an orientation introduces unfavorable steric interactions, and a  lower enthalpy can be obtained when the hydrogens point towards four cube faces instead. Because of this the four coordinate CH$_4$ species within the phase we refer to as $T_d$-CS$_{15}$H$_{48}$ actually possesses $C_{2v}$ symmetry. As illustrated in Figure\ \ref{fig:Td}a, at 270~GPa its two C-H bond lengths are nearly identical with calculated -iCOHPs of 6.34 and  6.30~eV/bond. The H-S 
distances between two of the hydrogens bonded to carbon elongate to 1.73~\AA{}, and a 
further two to 1.91~\AA{}. At the same time two of the S-H bonds contract relative to
those within $Im\bar{3}m$ H$_3$S (1.34~\AA{}). The encapsulated methane molecule 
possesses H-C-H angles that deviate from the ideal tetrahedral angle (100$^\circ$, 
103$^\circ$ and 140$^\circ$), and its Bader charge, -0.09, is suggestive of electron donation 
from the H-S lattice. Plots of the ELF (Figure\ \ref{fig:Td}b) clearly illustrate the C-H bond, but do 
not show any evidence of covalent bond formation between CH$_4$ and the H-S framework. Thus, 
two H-S and four C-H bonds per unit cell in $T_d$-CS$_{15}$H$_{48}$ become classical 2c-2e bonds, no longer participating  in delocalized multi-centered bonding as they would within H$_3$S. 

\subsection*{Properties of the S~$\rightarrow$~C Doped Phases at 270~GPa}
Our calculations find the DOS at the Fermi level, $N\textsubscript{F}$, in H$_3$S to be 0.050~states eV$^{-1}$\AA{}$^{-3}$ at 270 GPa. Moving $E\textsubscript{F}$ down in energy by 0.17~eV, which can be achieved by doping with $\sim$5.7\% C, yields the highest possible value of 0.055~states eV$^{-1}$\AA{}$^{-3}$. Given a reference material whose superconducting critical temperature, $T\textsubscript{c}^0$, is known, the approximate Bardeen-Cooper-Schrieffer formula can be employed to estimate the $T\textsubscript{c}$ of a similar material via the formula  $T\textsubscript{c} = 1.13\Theta\textsubscript{D}(T\textsubscript{c}^0/1.13\Theta\textsubscript{D}^0)^{N\textsubscript{F}^0/N\textsubscript{F}}$, where $\Theta\textsubscript{D}$ is the Debye 
temperature and $N\textsubscript{F}$ is given per unit volume \cite{Feng:2006a}. Using the measured value 
of $T\textsubscript{c}^0=$170~K at 270~GPa for $Im\bar{3}m$ H$_3$S \cite{Minkov:2020a}, this simple 
model predicts the $T\textsubscript{c}$ to increase to 208~K for $\sim$5.7\% C doping, which is somewhat 
lower than the results obtained using the VCA at the same pressure and doping level 
\cite{Ge:2020a}.

\begin{figure*}[b!]
\begin{center}
\includegraphics[width=\textwidth]{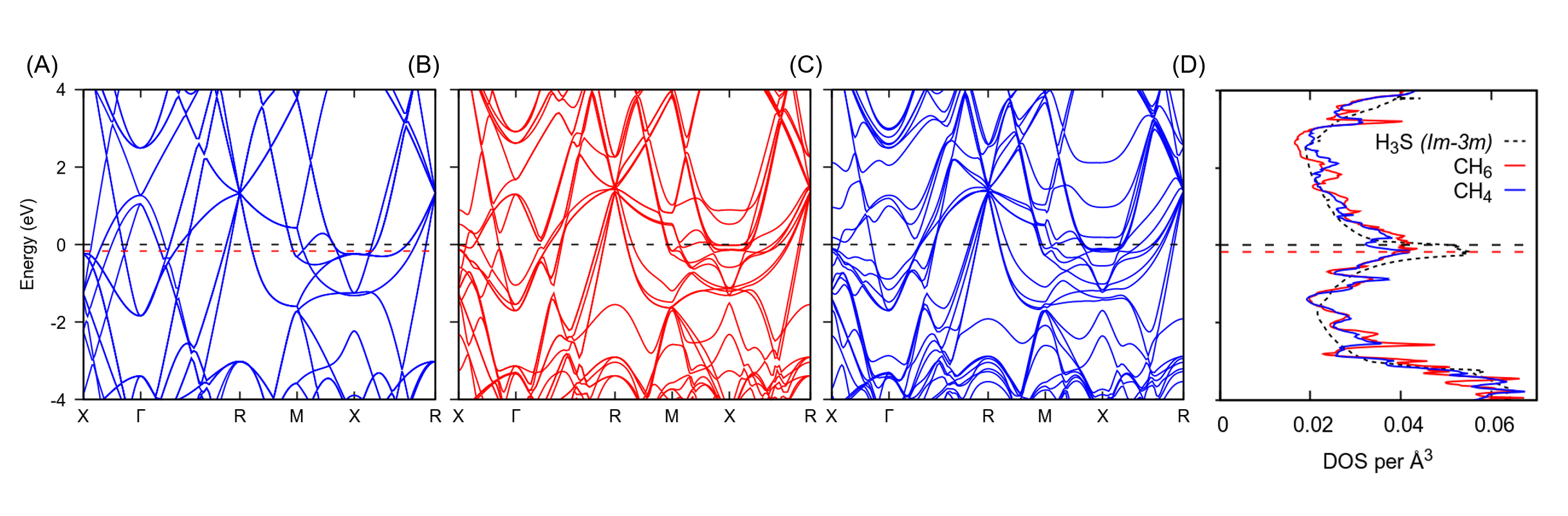}
\end{center}
\caption{Band structure at 270~GPa for (a) $Im\bar{3}m$ H$_3$S in a $2\times2\times2$ 
supercell of the standard conventional lattice, (b) $Pm\bar{3}m$ $O_h$-CS$_{15}$H$_{48}$, 
and (c) $Amm2$ $T_d$-CS$_{15}$H$_{48}$. To compare the band structures of the three 
phases, which all possess different space-groups, simple cubic symmetry is assumed and the 
high symmetry special points used are $\Gamma$ (0,0,0), $X$ (0.5,0,0), $M$ (0.5,0.5,0) and 
$R$ (0.5,0.5,0.5).  (d) Densities of states of these phases where CH$_6$ denotes 
CS$_{15}$H$_{48}$ with hexacoordinate carbon and CH$_4$ with tetracoordinate carbon. The black horizontal dashed line denotes $E\textsubscript{F}$ and the red dashed line $E\textsubscript{F}-
0.17$~eV, which corresponds to the top of the peak in the DOS in $Im\bar{3}m$ H$_3$S.
\label{fig:bands}}
\end{figure*}

To study the effect of doping on the electronic structure, we performed a single point calculation on an unrelaxed $2\times2\times2$ supercell of  $Im\bar{3}m$ H$_3$S where a single sulfur atom was replaced by carbon. In contrast to the VCA model predictions we found that $N\textsubscript{F}$ 
\emph{decreases} to 0.048~states eV$^{-1}$\AA{}$^{-3}$. Structural relaxation to the $O_h$- and $T_d$-CS$_{15}$H$_{48}$ phases further lowers $N\textsubscript{F}$ to 0.040 and 0.033~states eV$^{-1}$\AA{}$^{-3}$, respectively, as illustrated in Figure\ 
\ref{fig:bands}. This initially counter-intuitive behavior of $N\textsubscript{F}$  can be understood by considering the following: 
replacing sulfur by carbon followed by structural 
relaxation decreases the number of degenerate bands near $E\textsubscript{F}$ because some of the metallic 
electrons that were delocalized in the H$_3$S framework now become localized in covalent C-H 
bonds. Whereas 14 bands (some of which are degenerate) cross $E\textsubscript{F}-0.17$~eV in 
$Im\bar{3}m$ H$_3$S at 270~GPa, only 11 bands intersect with the Fermi level within both 
$O_h$- and $T_d$-CS$_{15}$H$_{48}$ (Figure\ \ref{fig:bands}a-\ref{fig:bands}c). 

We also investigated how the vibrational properties of $Im\bar{3}m$ H$_3$S are affected by 
carbon doping. The Debye temperature of the phase in 
which carbon is four-coordinate is larger than that containing six-coordinated carbon, which is 
higher than that of pure H$_3$S (Figure~S24). The frequencies of the asymmetric H-C-H stretching modes of 
$T_d$-CS$_{15}$H$_{48}$ were calculated to be 3125 and 3286~cm$^{-1}$, whereas the 
symmetric ones occurred at  3140 and 3173 cm$^{-1}$. $O_h$-CS$_{15}$H$_{48}$ possessed a 
single C-H stretching mode at 2358~cm$^{-1}$. Because the quasi-molecular CH$_4$ species 
has stronger and shorter C-H bonds as compared to the octahedrally coordinated carbon, these 
vibrations are found at higher frequencies. 

\subsection*{C$_x$S$_{1-x}$H$_3$ Phases as a Function of Doping and Pressure}
We now consider the dynamic stability of the C$_x$S$_{1-x}$H$_3$ stoichiometries as a 
function of doping at 270~GPa, beginning with phases where carbon is octahedrally 
coordinated. Both the H-S and H-C bond lengths in the 50\% doped dynamically unstable CSH$_6$ structure measured 1.37~\AA{}. A majority of the 
phonon modes that were imaginary at some point within the Brillouin Zone (BZ) turned out to 
be related to the motions of the hydrogen atoms bonded to carbon. Decreasing the doping 
level to 25\% allowed the C-H and S-H bonds to assume different values (1.08~\AA{} and 1.41-1.44~\AA{}, respectively). As a result, some of the imaginary modes in the 50\% doped 
structure become real, specifically the asymmetric and symmetric H-C-H stretching modes 
found near $\sim$2700 and 2500~cm$^{-1}$, respectively. Lowering the doping level left a 
single imaginary phonon mode throughout the whole BZ. This mode arose from the movement 
of a hydrogen atom sandwiched between two carbons that led to the lengthening of one, and 
contraction of another C-H bond that measured 1.43~\AA{} in the optimized geometry. The 
other C-H bond in this structure measured 1.13~\AA{}. Thus, the structural instability is 
a result of the too-long C-H bonds that arise because of the constraints imposed by the doping 
level and S-H lattice.

The $O_h$-CS$_{15}$H$_{48}$ phase possesses the stoichiometry with the lowest percentage 
doping that is a local minimum at 270~GPa. Phonon calculations showed 
this structure becomes dynamically unstable below 255~GPa (Figure~S4). Visualization of 
one of the triply degenerate imaginary modes at $\Gamma$ revealed that it can be described 
as a lengthening/contraction of the CH-S distance, which measured 1.73~\AA{} in the optimized structure, 
with the other two modes corresponding to the same vibration but along the other 
crystallographic axes. Upon decreasing pressure from 270 to 240~GPa the C-H bond length 
increased minimally ($\Delta d=$~0.005~\AA{}, -$\Delta$iCOHP~=~0.02~eV/bond), whereas the 
increase in the CH-S distance was an order of magnitude larger ($\Delta d=$~0.03~\AA{}, -$\Delta$iCOHP~=~0.10~eV/bond). These results suggest that the instability that emerges near 
255~GPa is primarily a result of decreased S-H interaction at lower pressures.

We also consider $O_h$-CS$_{53}$H$_{162}$, which corresponds to 1.85\% doping, with 
calculated C-H bond lengths of 1.16~\AA{} and S-H bonds ranging from 1.43-1.48~\AA{}. Unsurprisingly, its DOS at $E\textsubscript{F}$ of 0.0484~states eV$^{-
1}$\AA{}$^{-3}$ and Debye temperature of 1375~K approach the values obtained for 
$Im\bar{3}m$ H$_3$S. Phonon calculations revealed that this phase was dynamically stable at 270~GPa, and it could be stabilized to 
lower pressures than $O_h$-CS$_{15}$H$_{48}$, becoming unstable at approximately 160 GPa. 
The structures that possess four-coordinated carbon atoms are found to be dynamically stable 
at 270~GPa when the doping level was less than 25\% (CS$_3$H$_{12}$, CS$_7$H$_{24}$, 
CS$_{15}$H$_{48}$ and CS$_{53}$H$_{162}$). CS$_3$H$_{12}$ was unstable by 140~GPa and 
visualization of the largest magnitude imaginary mode found at the $R$ point illustrated that it 
corresponded to a symmetric/asymmetric H-S-H stretch. CS$_7$H$_{24}$ and 
CS$_{15}$H$_{48}$ became dynamically unstable near 250~GPa via a softening of the 
longitudinal acoustic mode at an off $\Gamma$ point, and $T_d$-CS$_{53}$H$_{162}$ was 
stable (unstable) at 200 (160)~GPa.

\subsection*{C$_x$S$_{1-x}$H$_{3+x}$ Phases: Doping H$_3$S via Substituting SH$_3$ by 
CH$_4$}

Instead of replacing a fraction of the S atoms by C atoms, another way to dope H$_3$S would 
be to replace some of the SH$_3$ units by CH$_4$ in a large supercell. Indeed, XRD 
and equation of state analysis show that the precursor phases of the C-S-H superconductor are 
(H$_2$S)$_2$H$_2$ and (CH$_4$)$_2$H$_2$ van der Waals compounds that have identical 
volumes at the synthesis pressure, thereby allowing readily mixed (H$_2$S,CH$_4$)$_2$H$_2$ 
alloys \cite{Lamichhane:2021}. This leads to the possibility that CH$_4$ molecules persist in the 
structure well into the superconducting H$_3$S-based phase or phases with H$_2$ taken up in the 
structure. First-principles calculations have previously been employed to investigate the 
properties of metastable phases that correspond to 50\% SH$_3\rightarrow$~CH$_4$ 
substitution, wherein methane molecules were intercalated in an H$_3$S framework.  
\cite{Zurek:2020b,Sun:2020a}. Lower dopings could be derived from the $T_d$-C$_x$S$_{1-
x}$H$_3$ phases discussed above by adding a single hydrogen atom to the S-H lattice. Unlike 
structures in which C replaced S, all of the CH$_4$ dopings we considered (25, 12.5, 6.25 and 
1.85\% C) were found to be dynamically stable at 270~GPa and remained so on decompression 
to at least 140~GPa (Figures S17, S19-S22).

The optimized geometry obtained by taking a sixteen formula unit supercell and replacing one 
SH$_3$ by CH$_4$, corresponding to a stoichiometry of CS$_{15}$H$_{49}$, is illustrated in Figure~S16c \cite{Zurek:2020b}. The Bader charge on the $C_{3v}$ symmetry methane molecule 
was nearly the same as in $T_d$-CS$_{15}$H$_{48}$, $-0.10$, but the bonds were somewhat 
shorter and stronger (1.03~\AA{} and 6.92~eV/bond ($\times$ 1), 1.04~\AA{} and 
6.98~eV/bond ($\times$ 3)) and the angles were closer to those of a perfect tetrahedron 
(105$^\circ$, 113$^\circ$). The DOS at $E\textsubscript{F}$ of CS$_{15}$H$_{49}$ (0.034~states eV$^{-
1}$\AA{}$^{-3}$) is quite comparable to that of $T_d$-CS$_{15}$H$_{48}$, but its Debye 
temperature is significantly higher (1825~K, Figure S24), suggesting that its $T\textsubscript{c}$ may be higher as well. 

\subsection*{Thermodynamic Properties and Equation of States}

\begin{figure*}
\begin{center}
\includegraphics[width=0.6\textwidth]{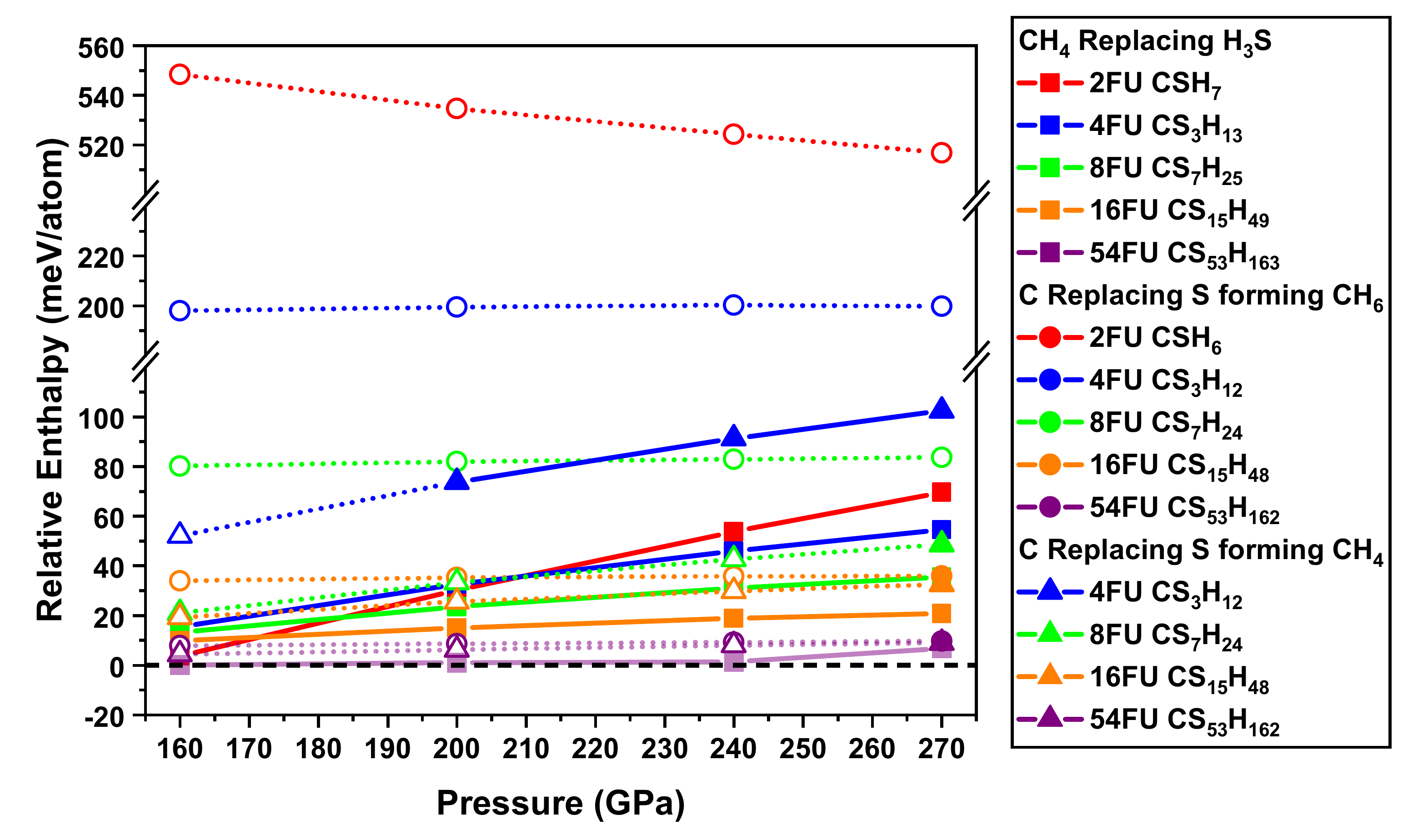}
\end{center}
\caption{$\Delta H$ as a function of pressure for the following reaction:  $ 
\textrm{C}+x(\textrm{H}_3\textrm{S})+\left(\frac{y-3x}{2}\right)\textrm{H}_2\rightarrow 
\textrm{CS}_x\textrm{H}_y$ in meV/atom using the diamond phase of carbon \cite{Martinez2012}, 
$C2/c$ phase of H$_2$ \cite{LiuMa:2012b} and $Im\bar{3}m$ phase of H$_3$S \cite{Li:2014} 
as a reference. The open (closed) circles correspond to structures that were found to be
dynamically unstable (stable).
\label{fig:enthalpy}}
\end{figure*}

The relative enthalpies of the doped structures from H$_3$S, as well as carbon in the diamond 
phase and the $C2/c$ phase of molecular H$_2$ (Figure\ \ref{fig:enthalpy}) illustrate that doping 
is thermodynamically unfavorable within the pressure range considered, consistent with 
previous studies of carbon doped SH$_3$ phases 
\cite{Zurek:2020b,Sun:2020a,Hu:2020,Wang:2021}. For a given number of C+S atoms the phase 
where CH$_4$ replaced H$_3$S was always the most stable, followed by phases where a 
tetrahedrally coordinated C replaced an S atom, and lastly those where C was octahedrally 
coordinated. Exploratory calculations suggested that it was not enthalpically favorable to add 
another hydrogen to the SH$_3\rightarrow$~CH$_4$ doped phases by forming another S-H 
bond (e.g.\ $\Delta H$ for the reaction 
$\textrm{CS}_3\textrm{H}_{13}+\frac{1}{2}\textrm{H}_2\rightarrow \textrm{CS}_3\textrm{H}_{14}$  was 
+17.8~meV/atom). 

The enthalpies of the phases with the lowest levels of doping were within $<2$~meV/atom of 
each other, and they were the closest to the threshold for thermodynamic stability, with 
$\Delta H$ for the formation of CS$_{53}$H$_{163}$ being 0.14 and 6.82~meV/atom at 140 
and 270~GPa, respectively. At 270~GPa the zero-point-energy disfavored the doped 
species where carbon was tetrahedrally coordinated (Table~S2) but this had less of an 
effect on the octahedrally coordinated systems, such that for S~$\rightarrow$~C replacements, 
the structures in which C was six coordinate became slightly preferred. Even though the carbon 
doped phases are not thermodynamically stable, their  enthalpies of formation are generally within the $\sim$70~meV/atom threshold corresponding to the 90$^\textrm{th}$ percentile of DFT-calculated metastability for inorganic crystalline materials at 1~atm \cite{materialsproject}.

\subsection*{Superconducting Properties}

In addition to the phonon band structures and phonon densities of states, we calculated the 
Eliashberg spectral function, $\alpha^2F(\omega)$, and the EPC integral, $\lambda(\omega)$, 
for the model structures considered here  (Figure~S1d, S7a, and S16a). Initially, $O_h$-CS$_{15}$H$_{48}$ was 
singled out for analysis because its high $N\textsubscript{F}$ relative to other carbon-doped phases 
suggests it may have the highest $T\textsubscript{c}$. Calculations for structures with lower carbon 
concentrations are currently computationally unfeasible because of their large supercells. Since 
$T\textsubscript{c}$ is usually highest near the onset of dynamic instability the calculation was performed at 
255~GPa (Figure S27). The estimated EPC (Table \ref{tab:tc}) was a modest $\lambda=$~1.09, 
with 27\% arising from the low frequency modes associated with the motions of the sulfur 
atoms ($<$700~cm$^{-1}$), and 72.5\% resulting from motions predominantly due 
to the hydrogen atoms bonded to sulfur, with a minor contribution from the 
hydrogens bonded to carbon (700-2200~cm$^{-1}$).  Numerical solution of the Eliashberg 
equations using a typical value of the Coulomb pseudopotential, $\mu^*=0.1$, resulted in a 
$T\textsubscript{c}$ of 136~K at 255~GPa.

The $T\textsubscript{c}$ computed for $O_h$-CS$_{15}$H$_{48}$ is over
100~K lower than the maximum measured for the C-S-H superconductor \cite{Snider:2020a}, in agreement with recent results of Guan \emph{et al.}\ \cite{Guan:2021}. Even though the Debye temperature of this phase is predicted to be higher than that of $Im\bar{3}m$ H$_3$S (at 270~GPa), its 
$\omega\textsubscript{ln}$ is lower, as is the $\lambda$ and resulting $T\textsubscript{c}$ predicted by the present approach. Decreasing the doping level will increase $T_c$. Nonetheless, our results suggest that the superconducting properties of structures belonging to this family are unlikely to surpass those of  $Im\bar{3}m$ H$_3$S because $N\textsubscript{F}$ of the doped phase is lower, but the Debye temperatures are similar (cf.\ values obtained at 270~GPa in Table \ref{tab:tc}).

\begin{table*}
    \centering
    \caption{Debye temperature ($\Theta\textsubscript{D}$), density of states at the Fermi level ($N\textsubscript{F}$, in states/eV/\AA$^3$), logarithmic average of phonon frequencies ($\omega\textsubscript{ln}$), electron-phonon coupling constant ($\lambda$), and superconducting critical temperature ($T\textsubscript{c}$)  obtained via numerical solution of the Eliashberg equations using $\mu^*=0.1$ at 270~GPa.  The pressure range where dynamic stability was confirmed in this study (phonon calculations were carried out between 140-270~GPa) is also provided. }
    \begin{tabular}{lcccccc}
        \hline
        Structure & $\Theta\textsubscript{D}$ (K) & $N\textsubscript{F}$  & $\omega\textsubscript{ln}$ (K) & $\lambda$ & $T\textsubscript{c}$ (K) & $P\textsubscript{stab}$ (GPa)\\
        \hline
        $Im\bar{3}m$ H$_3$S  & 1314 & 0.0502 & 1603 & 1.25 & 174 & N/A\\
        $R3m$ H$_3$S & 1340 & 0.0440 & 1615 & 0.95 & 140 & N/A\\
        \hline
        $^aO_h$-CS$_{15}$H$_{48}$ & 1449 & 0.0401 & & & & 260-270\\
        $O_h$-CS$_{53}$H$_{162}$ & 1375 & 0.0484 & & & & 200-270 \\
        \hline
        $T_d$-CS$_3$H$_{12}$ & 1813 & 0.0281 & 1370 & 0.85 & 80 & 180-270 \\
        $T_d$-CS$_7$H$_{24}$ & 1875 & 0.0395 &  & & & 270 \\
        $T_d$-CS$_{15}$H$_{48}$ & 1639 & 0.0325 & & & & 270 \\
        $T_d$-CS$_{53}$H$_{162}$ & 1774 & 0.0460 & & & & 260-270\\
        \hline
        $Cmcm$ CSH$_7$ & 1781 & 0.0256 & & & & 140-270\\
        $R3m$ CSH$_7$ & 1851 & 0.0275 & & & & 140-270\\
        CS$_3$H$_{13}$ & 1821 & 0.0245 & 1704 & 1.01 & 142 & 140-270 \\
        CS$_7$H$_{25}$ & 1868 & 0.0313 & & & & 140-270\\
        CS$_{15}$H$_{49}$ & 1825 & 0.0343 & & & & 140-270\\
        $^b$CS$_{53}$H$_{163}$ & 1818 & 0.0434 & & & & 140-270 \\
        \hline
    \end{tabular} \\
    $^a$ The computed values for this phase at 255~GPa are: 
    $\Theta\textsubscript{D}=$1490~K,  $N\textsubscript{F}=$1414~K, $\lambda=$1.09, $T\textsubscript{c}=$136~K (numerically solving Eliashberg equations). \\
    $^b$ Assuming $\omega\textsubscript{ln}=$1704~K and $\lambda=$2.5 yields $T\textsubscript{c}=$280~K (calculated using the Allen-Dynes modified McMillan equation -- see text).
    \label{tab:tc}
\end{table*}

We now examine whether tetrahedral coordination of the carbon atoms can enhance $T\textsubscript{c}$. 
Unfortunately, because the symmetries of $T_d$-CS$_{15}$H$_{48}$ and CS$_{15}$H$_{49}$ 
are lower than that of $O_h$-CS$_{15}$H$_{48}$ calculating their $T\textsubscript{c}$s was prohibitively 
expensive. Therefore, 25\% doped systems were considered instead. Despite the fact that the 
estimated Debye temperature for $T_d$-CS$_{15}$H$_{48}$ was higher than that of H$_3$S, 
$\omega\textsubscript{ln}$ was lower, mirroring the findings for $O_h$-CS$_{15}$H$_{48}$. In 
addition, the EPC was lower, consistent with the much smaller $N\textsubscript{F}$, resulting in a $T\textsubscript{c}$ of 
80~K. Even though decreasing the doping increases $N\textsubscript{F}$, once again it is unlikely that 
compounds belonging to this family of structures could be superconducting at temperatures 
higher than those found for H$_3$S. However, adding one more hydrogen atom per unit cell to 
members of this family leads to a remarkable improvement. Both $\omega\textsubscript{ln}$ and 
$\lambda$ of CS$_3$H$_{13}$ are higher than that of $T_d$-CS$_3$H$_{12}$, yielding a $T\textsubscript{c}$ 
of 142~K at 270~GPa. Thus, a little bit of carbon does a lot for superconductivity in H$_3$S, and the way in which it is altered (whether it be an increase or decrease) is intimately related to the local bonding configuration around carbon.

To better understand how adding a single hydrogen atom can dramatically increase $T\textsubscript{c}$ by 
60~K, we compare $\lambda(\omega)$ and $\omega\textsubscript{ln}(\omega)$ of these two model 
structures (Figure~S24). At $\sim$1550~cm$^{-1}$ their $\omega\textsubscript{ln}(\omega)$ are 
almost identical. At higher frequencies $\omega\textsubscript{ln}(\omega)$ for CS$_3$H$_{13}$ 
increases much faster until a near plateau region is attained at 2220~cm$^{-1}$ (1670~K). In 
$T_d$-CS$_3$H$_{12}$, on the other hand, $\omega\textsubscript{ln}(\omega)$ reaches 1309 K at 
2450~cm$^{-1}$. Beyond $\sim$2500~cm$^{-1}$, the flat high-frequency bands contribute less 
than 40~K to $\omega\textsubscript{ln}(\omega)$ for both structures. Therefore, the main difference 
in the total $\omega\textsubscript{ln}$ for the two phases arises from modes in the 1550-2500~cm$^{-
1}$ frequency range. From the projected phonon density of states (Figure~S10a and S17a) 
modes associated with motions of the H atoms in the H-S lattice (H1) primarily comprise this 
region, with some contribution from H atoms in the CH$_4$ molecule (H2). At 
$\sim$1410~cm$^{-1}$ $\lambda(\omega)$ was almost the same for the two structures 
($\sim$0.45). At higher frequencies the $\lambda(\omega)$ for CS$_3$H$_{13}$ increases 
much faster until 2220~cm$^{-1}$ where it reaches a value of 1.00, while for $T_d$-CS$_3$H$_{12}$ $\lambda(\omega)$ reaches 0.83 at 2450~cm$^{-1}$. Beyond 
$\sim$2500~cm$^{-1}$, the high-frequency flat bands contribute less than 0.005 towards 
$\lambda$ for both structures.

\begin{figure*}
\begin{center}
\includegraphics[width=0.6\textwidth]{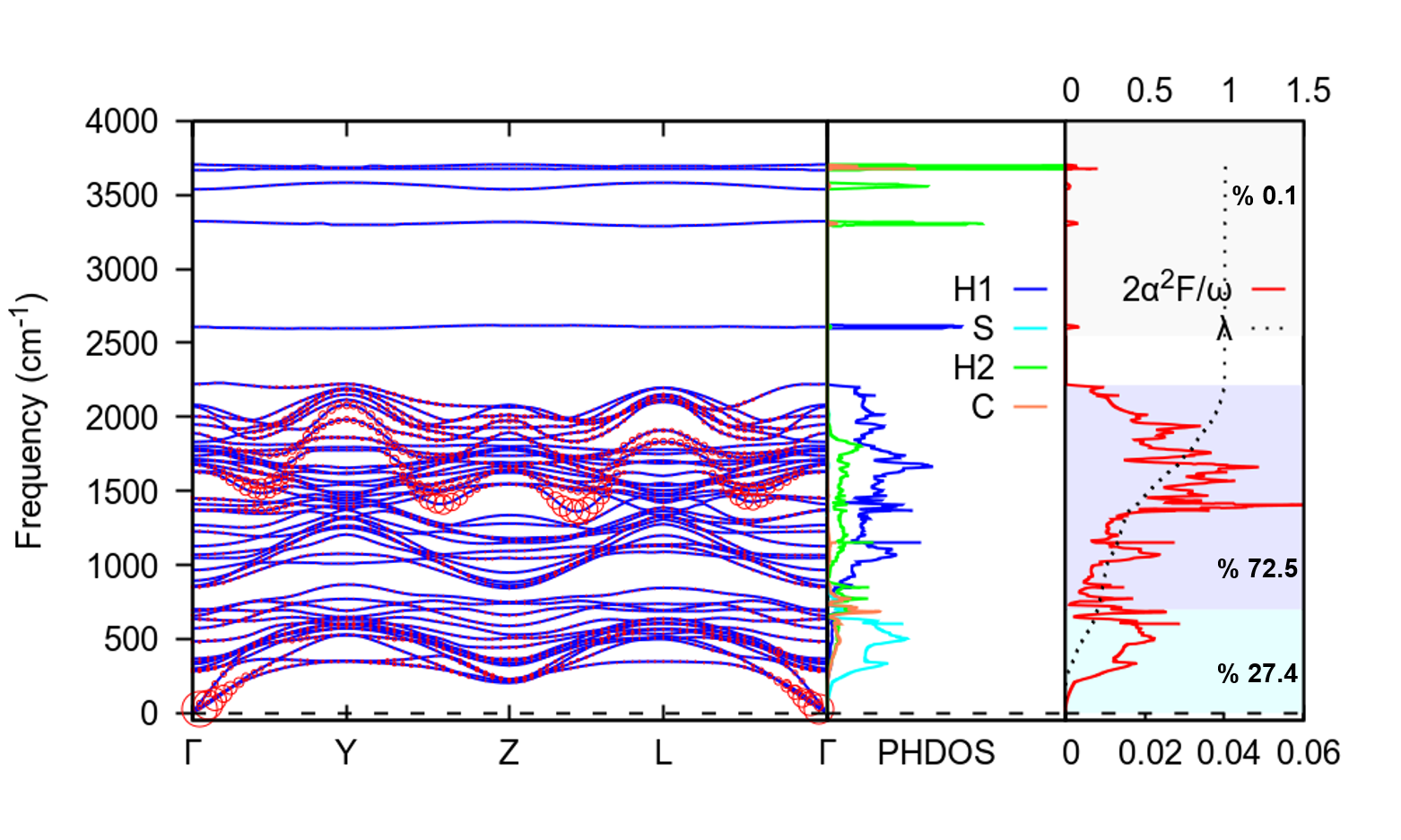}
\end{center}
\caption{Phonon band structure, phonon density of states (PHDOS), Eliashberg spectral 
function in the form of $\frac{2\alpha ^2F(\omega)}{\omega}$, and the electron phonon
integral, $\lambda(\omega)$, for CS$_{3}$H$_{13}$ at 270~GPa. Red circles indicate the 
electron-phonon coupling constant, $\lambda_{q\nu}$, at mode $\nu$ and wavevector $q$, 
and their radii is proportional to the strength. H1 are the hydrogen atoms in the S-H lattice, and 
H2 are the hydrogen atoms in CH$_4$. The percentages given were calculated via 
$\left(\int_{\omega_1}^{\omega_2}\lambda(\omega)d\omega / \lambda\right)\times 100\%$; 
and the frequency regions spanning $\omega_1$ and $\omega_2$ are colored coded in the 
PHDOS.
\label{fig:epc}}
\end{figure*}

Additional insight into the types of motions that result in a larger $\lambda$ within 
CS$_3$H$_{13}$ can be obtained by visualizing selected vibrational modes that have a notable 
$\lambda_{\textrm{\textbf{q}}\nu}$, e.g., modes about halfway along the Y$\rightarrow$Z 
(1412~cm$^{-1}$) and Z$\rightarrow$L (1361 cm$^{-1}$) paths, which have very large contributions 
(Figure\ \ref{fig:epc}). Both of these involve H2 vibrations between neighboring S 
atoms, leading to a snakelike undulation of the H1/H2 chains. The addition of the extra S-bonded H atom expands the S-H lattice in the $ab$ plane, leaving looser contacts between the 
S and H atoms (1.44~\AA{} and 1.43~\AA{} as opposed to 1.41-1.43~\AA{} in $T_d$-CS$_3$H$_{12}$), thereby softening the phonon frequencies. The $c$ axis is largely 
unchanged by the addition of an extra H, but the H-S distances that are relatively even in CS$_3$H$_{12}$ (1.43-1.44~\AA{}) become disproportionate in CS$_3$H$_{13}$ to 1.41 and 1.46~\AA{}.  Within the 
methane fragments, two H1 atoms possessed mirrored circular motions that were part of the 
undulating H chains.

Although decreasing the doping does not appreciably change the Debye temperature, it 
increases the $N\textsubscript{F}$ substantially such that at 1.85\% doping the value is almost double that 
of the 25\% doped phase. Assuming that $\omega\textsubscript{ln}$ for CS$_{53}$H$_{163}$ is close 
to that for CS$_3$H$_{13}$ a $T\textsubscript{c}$ of 280~K could be obtained using the Allen-Dynes modified 
McMillan equation \cite{Allen:1975} for $\lambda\sim2.5$. Since this approximate equation is 
known to underestimate $T\textsubscript{c}$s for systems where $\lambda\ge$1.5, within the Eliashberg 
formalism somewhat smaller $\lambda$ values would yield a similar result. Therefore, it is 
plausible that the $\lambda$ of CS$_{53}$H$_{163}$ could fall within this range given its high 
$N\textsubscript{F}$. For comparison, at 200~GPa $Im\bar{3}m$ H$_3$S is computed to have a similar
$\lambda$ (2.19), but its  $\omega\textsubscript{ln}$ (1335~K) is significantly smaller, yielding an 
estimated $T\textsubscript{c}$ of 191-204~K \cite{Duan:2014}. Moreover, its DOS at $E\textsubscript{F}$, 0.041 
states/eV/\AA$^3$ , is comparable to that obtained for CS$_{53}$H$_{163}$. It is beyond the 
scope of this study to explicitly calculate the $T\textsubscript{c}$ of C$_x$S$_{1-x}$H$_{3+x}$ phases over a 
broad range of doping levels (e.g., 1.85-12.5\%). We note that unlike the other 
families of structures studied to date, structures belonging to this family  were found to be 
dynamically stable across a broad pressure range (at least 140-270~GPa).  

%
In summary, our detailed computations have shown that doping H$_3$S by 1.85-25\% of carbon at 270 GPa leads to a plethora of metastable phases where carbon can 
be either six-coordinated or four-coordinated to hydrogen. In the first case we see the 
remarkable emergence of an $O_h$ symmetry CH$_6$ motif reminiscent of a di-carbocation, but 
stabilized in the solid state under pressure via weak interactions by the negatively charged 
environment of the surrounding host H$_3$S-like lattice. The second case is an example of 
methane intercalated within an H$_3$S-like framework. These doping schemes split 
degenerate bands thereby decreasing $N_F$, and  localizing electrons in covalent C-H 
bonds whose signatures are far removed from the Fermi energy. 

Both $\omega\textsubscript{ln}$ and $\lambda$ of 
$O_h$-CS$_{15}$H$_{48}$ and $T_d$-CS$_{3}$H$_{12}$ were smaller than the values 
computed for $Im\bar{3}m$ H$_3$S, as was the resulting $T\textsubscript{c}$. Remarkably, adding a single hydrogen atom to $T_d$-CS$_{3}$H$_{12}$ increased $T\textsubscript{c}$ by 60~K. The larger 
$\lambda$ and $\omega\textsubscript{ln}$ of the resulting CS$_3$H$_{13}$ phase could be traced 
back to the emergence of soft phonon modes, a consequence of the weaker and longer S-H 
bonds in the host lattice caused by the insertion of the extra S-bonded hydrogen atom. At 
270~GPa the $T\textsubscript{c}$ of CS$_3$H$_{13}$ was about the same as that of $R3m$ H$_3$S despite 
its much lower DOS at $E\textsubscript{F}$.  Decreasing the doping level in this family of structures, derived from H$_3$S by low-level incorporation of CH$_4$, is likely to increase $T_c$ further.
We hope that the present results will stimulate further theoretical 
studies, large supercell calculations of the superconducting properties of these structures, as 
well as additional experiments in search of still higher $T\textsubscript{c}$s and phases that 
may be stable over a broader range of conditions. 

\section*{Methods}
\subsection*{Electronic Structure Calculations} Geometry optimizations and electronic structure calculations were performed using density 
functional theory (DFT) with the Perdew-Burke-Ernzerhof (PBE) exchange correlation functional 
\cite{perdew1996generalized} as implemented in the Vienna \textit{Ab~initio} Simulation 
Package (VASP) 5.4.1 \cite{kresse1996efficient}. The valence electrons (H 
1s$^1$, S 3s$^2$3p$^4$, and C 2s$^2$p$^2$) were treated explicitly using plane wave basis 
sets with a cutoff energy of 800~eV (as previously employed by Cui \textit{et 
al.}\cite{Zurek:2020b}), while the core states were treated with the projector-augmented wave 
(PAW) \cite{blochl1994projector} method. The reciprocal space was sampled using a 
$\Gamma$-centered Monkhorst-Pack scheme, and the number of 
divisions along each reciprocal lattice vector was chosen such that the product of this number 
with the real lattice constant was 70~\AA{} for the density of states (DOS) calculations, and 
50~\AA{} otherwise. The crystal orbital Hamilton populations (COHPs) 
\cite{Dronskowski:1993}, the negative of the COHPs integrated to the Fermi level (-iCOHPs), and the crystal orbital bond index (COBI) \cite{Muller:2021} were calculated using the  LOBSTER package (v2.2.1 for -iCOHP and v4.1.0 for COBI) \cite{Maintz:2013}, and the 
results used to analyze the bonding. The dynamic stability of the phases was investigated via 
phonon calculations performed using the finite difference scheme as implemented in the 
PHONOPY software package \cite{togo2008first}.
For the DFT-raMO procedure \cite{DFTraMO}, single point calculations were carried out in VASP with coarse $k$-point grids sampling the full Brillouin zone with dimensions equal to the supercell used in the raMO analysis, corresponding to 3$\times$3$\times$3 for CS$_{15}$H$_{48}$ and 5$\times$5$\times$5 for H$_3$S.

\subsection*{Superconducting Properties} Phonon calculations were performed using the Quantum Espresso (QE) \cite{giannozzi2009quantum} package program to obtain the dynamical matrix and the 
electron-phonon coupling (EPC) parameters. The pseudopotentials were obtained from the  
PSlibrary \cite{dal2014pseudopotentials} using H $1s^1$, S $3s^23p^4$, and C $2s^22p^2$ 
valence electrons and the PBE exchange-correlation functional \cite{perdew1996generalized}. 
Plane-wave basis set cutoff energies were set to 80~Ry for all systems. We employed a $\Gamma$-centered Monkhorst-Pack Brillouin zone sampling scheme, along with Methfessel-Paxton smearing with a broadening width of 0.02~Ry.  Density functional perturbation theory (DFPT) as implemented in QE was employed for the phonon calculations.  The EPC matrix elements were calculated using the $k$ and $q$-meshes, and Gaussian broadenings listed in Table~S1. The EPC parameter ($\lambda$) converges to within 0.05 for differences of the Gaussian broadening that are less than 0.02~Ry. The critical superconducting temperature, $T\textsubscript{c}$, has been estimated using the Allen-Dynes 
modified McMillan equation \cite{Allen:1975}, $
    T\textsubscript{c} = \frac{\omega\textsubscript{ln}}{1.2}\exp\left[-\frac{1.04(1+\lambda)}{\lambda-
\mu^*(1+0.62\lambda)}\right]$,
where $\omega\textsubscript{ln}$ is the logarithmic average frequency, and $\mu^*$ was set to 0.1.
The $T\textsubscript{c}$ were also obtained by solving the Eliashberg equations \cite{eliashberg:1960} 
numerically based on the spectral function, $\alpha^2F(\omega)$, obtained from the QE 
calculations.


\begin{thebibliography}{10}
\expandafter\ifx\csname url\endcsname\relax
  \def\url#1{\burl{#1}}\fi
\expandafter\ifx\csname urlprefix\endcsname\relax\def\urlprefix{URL }\fi
\providecommand{\bibinfo}[2]{#2}
\providecommand{\eprint}[2][]{\url{#2}}
\providecommand{\doi}[1]{\url{https://doi.org/#1}}

\bibitem{Ashcroft:2004a}
\bibinfo{author}{Ashcroft, N.~W.}
\newblock \bibinfo{title}{Hydrogen dominant metallic alloys: High temperature
  superconductors?}
\newblock \emph{\bibinfo{journal}{Phys. Rev. Lett.}}
  \textbf{\bibinfo{volume}{92}}, \bibinfo{pages}{187002 (1--4)}
  (\bibinfo{year}{2004})

\bibitem{Zurek:2009c}
\bibinfo{author}{Zurek, E.}, \bibinfo{author}{Hoffmann, R.},
  \bibinfo{author}{Ashcroft, N.~W.}, \bibinfo{author}{Oganov, A.~R.} \&
  \bibinfo{author}{Lyakhov, A.~O.}
\newblock \bibinfo{title}{A little bit of lithium does a lot for hydrogen}.
\newblock \emph{\bibinfo{journal}{Proc. Natl. Acad. Sci.}}
  \textbf{\bibinfo{volume}{106}}, \bibinfo{pages}{17640--17643}
  (\bibinfo{year}{2009})

\bibitem{Zurek:2018m}
\bibinfo{author}{Zurek, E.} \& \bibinfo{author}{Bi, T.}
\newblock \bibinfo{title}{High-temperature superconductivity in alkaline and
  rare earth polyhydrides at high pressure: A theoretical perspective}.
\newblock \emph{\bibinfo{journal}{J. Chem. Phys.}}
  \textbf{\bibinfo{volume}{150}}, \bibinfo{pages}{050901 (1--13)}
  (\bibinfo{year}{2019})

\bibitem{Zurek:2018d}
\bibinfo{author}{Bi, T.}, \bibinfo{author}{Zarifi, N.},
  \bibinfo{author}{Terpstra, T.} \& \bibinfo{author}{Zurek, E.}
\newblock \bibinfo{title}{ in \textit{The search for superconductivity in high
  pressure hydrides}} (ed.\bibinfo{editor}{Reedijk, J.})
  \emph{\bibinfo{booktitle}{Elsevier Reference Module in Chemistry, Molecular
  Sciences and Chemical Engineering}} \bibinfo{pages}{1--36}
  (\bibinfo{publisher}{Elsevier}, \bibinfo{address}{Waltham, MA},
  \bibinfo{year}{2019})

\bibitem{Livas:2021a}
\bibinfo{author}{Flores-Livas, J.~A.} \emph{et~al.}
\newblock \bibinfo{title}{A perspective on conventional high-temperature
  superconductors at high pressure: Methods and materials}.
\newblock \emph{\bibinfo{journal}{Phys. Rep.}} \textbf{\bibinfo{volume}{856}},
  \bibinfo{pages}{1--78} (\bibinfo{year}{2020})

\bibitem{Drozdov:2015a}
\bibinfo{author}{Drozdov, A.~P.}, \bibinfo{author}{Eremets, M.~I.},
  \bibinfo{author}{Troyan, I.~A.}, \bibinfo{author}{Ksenofontov, V.} \&
  \bibinfo{author}{Shylin, S.~I.}
\newblock \bibinfo{title}{Conventional superconductivity at 203 kelvin at high
  pressures in the sulfur hydride system}.
\newblock \emph{\bibinfo{journal}{Nature}} \textbf{\bibinfo{volume}{525}},
  \bibinfo{pages}{73--76} (\bibinfo{year}{2015})

\bibitem{Somayazulu:2019-La}
\bibinfo{author}{Somayazulu, M.} \emph{et~al.}
\newblock \bibinfo{title}{Evidence for superconductivity above 260 {K} in
  lanthanum superhydride at megabar pressures}.
\newblock \emph{\bibinfo{journal}{Phys. Rev. Lett.}}
  \textbf{\bibinfo{volume}{122}}, \bibinfo{pages}{027001}
  (\bibinfo{year}{2019})

\bibitem{Drozdov:2019-La}
\bibinfo{author}{Drozdov, A.~P.} \emph{et~al.}
\newblock \bibinfo{title}{Superconductivity at 250 {K} in lanthanum hydride
  under high pressures}.
\newblock \emph{\bibinfo{journal}{Nature}} \textbf{\bibinfo{volume}{569}},
  \bibinfo{pages}{528--531} (\bibinfo{year}{2019})

\bibitem{Snider:2020a}
\bibinfo{author}{Snider, E.} \emph{et~al.}
\newblock \bibinfo{title}{Room-temperature superconductivity in a carbonaceous
  sulfur hydride}.
\newblock \emph{\bibinfo{journal}{Nature}} \textbf{\bibinfo{volume}{586}},
  \bibinfo{pages}{373--377} (\bibinfo{year}{2020})

\bibitem{Bykova:2021}
\bibinfo{author}{Bykova, E.} \emph{et~al.}
\newblock \bibinfo{title}{Structure and composition of {C-S-H} compounds up to
  143 {GPa}}.
\newblock \emph{\bibinfo{journal}{Phys. Rev. B}}
  \textbf{\bibinfo{volume}{103}}, \bibinfo{pages}{L14015}
  (\bibinfo{year}{2021})

\bibitem{Lamichhane:2021}
\bibinfo{author}{Lamichhane, A.} \emph{et~al.}
\newblock \bibinfo{title}{X-ray diffraction and equation of state of the
  {C-S-H} room-temperature superconductor}.
\newblock \emph{\bibinfo{journal}{J. Chem. Phys.}}
  \textbf{\bibinfo{volume}{155}}, \bibinfo{pages}{1} (\bibinfo{year}{2021})

\bibitem{Yao-S-review:2018}
\bibinfo{author}{Yao, Y.} \& \bibinfo{author}{Tse, J.~S.}
\newblock \bibinfo{title}{Superconducting hydrogen sulfide}.
\newblock \emph{\bibinfo{journal}{Chem. Eur. J}} \textbf{\bibinfo{volume}{24}},
  \bibinfo{pages}{1769--1778} (\bibinfo{year}{2017})

\bibitem{Strobel:2011a}
\bibinfo{author}{Strobel, T.~A.}, \bibinfo{author}{Ganesh, P.},
  \bibinfo{author}{Somayazulu, M.}, \bibinfo{author}{Kent, P. R.~C.} \&
  \bibinfo{author}{Hemley, R.~J.}
\newblock \bibinfo{title}{Novel cooperative interactions and structural
  ordering in {H$_2$S-H$_2$}}.
\newblock \emph{\bibinfo{journal}{Phys. Rev. Lett.}}
  \textbf{\bibinfo{volume}{107}}, \bibinfo{pages}{255503 (1--4)}
  (\bibinfo{year}{2011})

\bibitem{Li:2014}
\bibinfo{author}{Li, Y.}, \bibinfo{author}{Hao, J.}, \bibinfo{author}{Liu, H.},
  \bibinfo{author}{Li, Y.} \& \bibinfo{author}{Ma, Y.}
\newblock \bibinfo{title}{The metallization and superconductivity of dense
  hydrogen sulfide}.
\newblock \emph{\bibinfo{journal}{J. Chem. Phys.}}
  \textbf{\bibinfo{volume}{140}}~(17), \bibinfo{pages}{174712 (1--7)}
  (\bibinfo{year}{2014})

\bibitem{Duan:2014}
\bibinfo{author}{Duan, D.} \emph{et~al.}
\newblock \bibinfo{title}{Pressure-induced metallization of dense
  {(H$_2$S)$_2$H$_2$} with high-{T$_c$} superconductivity}.
\newblock \emph{\bibinfo{journal}{Sci. Rep.}} \textbf{\bibinfo{volume}{4}},
  \bibinfo{pages}{6968 (1--6)} (\bibinfo{year}{2014})

\bibitem{Einaga:2016}
\bibinfo{author}{Einaga, M.} \emph{et~al.}
\newblock \bibinfo{title}{Crystal structure of 200 {K}-superconducting phase of
  sulfur hydride system}.
\newblock \emph{\bibinfo{journal}{Nat. Phys.}} \textbf{\bibinfo{volume}{12}},
  \bibinfo{pages}{835--838} (\bibinfo{year}{2016})

\bibitem{Minkov:2020a}
\bibinfo{author}{Minkov, V.~S.}, \bibinfo{author}{Prakapenka, V.~B.},
  \bibinfo{author}{Greenberg, E.} \& \bibinfo{author}{Eremets, M.~I.}
\newblock \bibinfo{title}{A boosted critical temperature of 166 {K} in
  superconducting {D$_3$S} synthesized from elemental sulfur and hydrogen}.
\newblock \emph{\bibinfo{journal}{Angew. Chem. Int. Ed.}}
  \textbf{\bibinfo{volume}{59}}, \bibinfo{pages}{38970--18974}
  (\bibinfo{year}{2020})

\bibitem{Guigue:2017}
\bibinfo{author}{Guigue, B.}, \bibinfo{author}{Marizy, A.} \&
  \bibinfo{author}{Loubeyre, P.}
\newblock \bibinfo{title}{Direct synthesis of pure {H$_3$S} from {S} and {H}
  elements: No evidence of the cubic superconducting phase up to 160 {GPa}}.
\newblock \emph{\bibinfo{journal}{Phys. Rev. B}}
  \textbf{\bibinfo{volume}{95}}~(2), \bibinfo{pages}{020104(R) (1--5)}
  (\bibinfo{year}{2017})

\bibitem{Goncharov:2016a}
\bibinfo{author}{Goncharov, A.~F.} \emph{et~al.}
\newblock \bibinfo{title}{Hydrogen sulfide at high pressure: Change in
  stoichiometry}.
\newblock \emph{\bibinfo{journal}{Phys. Rev. B}} \textbf{\bibinfo{volume}{93}},
  \bibinfo{pages}{174105 (1--7)} (\bibinfo{year}{2016})

\bibitem{Laniel:2020a}
\bibinfo{author}{Laniel, D.} \emph{et~al.}
\newblock \bibinfo{title}{Novel sulfur hydrides synthesized at extreme
  conditions}.
\newblock \emph{\bibinfo{journal}{Phys. Rev. B}}
  \textbf{\bibinfo{volume}{102}}, \bibinfo{pages}{134109}
  (\bibinfo{year}{2020})

\bibitem{Akashi:2015-S}
\bibinfo{author}{Akashi, R.}, \bibinfo{author}{Kawamura, M.},
  \bibinfo{author}{Tsuneyuki, S.}, \bibinfo{author}{Nomura, Y.} \&
  \bibinfo{author}{Arita, R.}
\newblock \bibinfo{title}{First-principles study of the pressure and
  crystal-structure dependences of the superconducting transition temperature
  in compressed sulfur hydrides}.
\newblock \emph{\bibinfo{journal}{Phys. Rev. B}}
  \textbf{\bibinfo{volume}{91}}~(22), \bibinfo{pages}{224513 (1--7)}
  (\bibinfo{year}{2015})

\bibitem{Errea:2015a}
\bibinfo{author}{Errea, I.} \emph{et~al.}
\newblock \bibinfo{title}{Hydrogen sulphide at high pressure: A
  strongly-anharmonic phonon-mediated superconductor}.
\newblock \emph{\bibinfo{journal}{Phys. Rev. Lett.}}
  \textbf{\bibinfo{volume}{114}}, \bibinfo{pages}{157004 (1--5)}
  (\bibinfo{year}{2015})

\bibitem{Li:2016-S}
\bibinfo{author}{Li, Y.} \emph{et~al.}
\newblock \bibinfo{title}{Dissociation products and structures of solid
  {H$_2$S} at strong compression}.
\newblock \emph{\bibinfo{journal}{Phys. Rev. B}}
  \textbf{\bibinfo{volume}{93}}~(2), \bibinfo{pages}{020103(R) (1--5)}
  (\bibinfo{year}{2016})

\bibitem{Ishikawa:2016}
\bibinfo{author}{Ishikawa, T.} \emph{et~al.}
\newblock \bibinfo{title}{Superconducting {H$_5$S$_2$} phase in sulfur-hydrogen
  system under high-pressure}.
\newblock \emph{\bibinfo{journal}{Sci. Rep.}} \textbf{\bibinfo{volume}{6}},
  \bibinfo{pages}{23160 (1--8)} (\bibinfo{year}{2016})

\bibitem{Akashi:2016a}
\bibinfo{author}{Akashi, R.}, \bibinfo{author}{Sano, W.},
  \bibinfo{author}{Arita, R.} \& \bibinfo{author}{Tsuneyuki, S.}
\newblock \bibinfo{title}{Possible ``magneli'' phases and self-alloying in the
  superconducting sulfur hydride}.
\newblock \emph{\bibinfo{journal}{Phys. Rev. Lett.}}
  \textbf{\bibinfo{volume}{117}}, \bibinfo{pages}{075503 (1--6)}
  (\bibinfo{year}{2016})

\bibitem{Gordon:2016}
\bibinfo{author}{Gordon, E.~E.} \emph{et~al.}
\newblock \bibinfo{title}{Structure and composition of the
  200~k-superconducting phase of {H$_2$S} at ultrahigh pressure: The perovskite
  {(SH$^-$)(H$_3$S$^+$)}}.
\newblock \emph{\bibinfo{journal}{Angew. Chem. Int. Ed.}}
  \textbf{\bibinfo{volume}{55}}~(11), \bibinfo{pages}{3682--3684}
  (\bibinfo{year}{2016})

\bibitem{Majumdar:S-2017}
\bibinfo{author}{Majumdar, A.}, \bibinfo{author}{Tse, J.~S.} \&
  \bibinfo{author}{Yao, Y.}
\newblock \bibinfo{title}{Modulated structure calculated for superconducting
  hydrogen sulfide}.
\newblock \emph{\bibinfo{journal}{Angew. Chem. Int. Ed.}}
  \textbf{\bibinfo{volume}{56}}, \bibinfo{pages}{11390--11393}
  (\bibinfo{year}{2017})

\bibitem{Majumdar:S-2019}
\bibinfo{author}{Majumdar, A.}, \bibinfo{author}{Tse, J.~S.} \&
  \bibinfo{author}{Yao, Y.}
\newblock \bibinfo{title}{Mechanism for the structural transformation to the
  modulated superconducting phase of compressed hydrogen sulfide}.
\newblock \emph{\bibinfo{journal}{Sci. Rep.}} \textbf{\bibinfo{volume}{9}},
  \bibinfo{pages}{5023} (\bibinfo{year}{2019})

\bibitem{Verma:2018a}
\bibinfo{author}{Verma, A.~K.} \& \bibinfo{author}{Modak, P.}
\newblock \bibinfo{title}{A unique metallic phase of {H$_3$S} at high-pressure:
  Sulfur in three different local environments}.
\newblock \emph{\bibinfo{journal}{Phys. Chem. Chem. Phys.}}
  \textbf{\bibinfo{volume}{20}}, \bibinfo{pages}{26344--26350}
  (\bibinfo{year}{2018})

\bibitem{Errea:2016}
\bibinfo{author}{Errea, I.} \emph{et~al.}
\newblock \bibinfo{title}{Quantum hydrogen-bond symmetrization in the
  superconducting hydrogen sulfide system}.
\newblock \emph{\bibinfo{journal}{Nature}} \textbf{\bibinfo{volume}{532}},
  \bibinfo{pages}{81--84} (\bibinfo{year}{2016})

\bibitem{Bianco:2018a}
\bibinfo{author}{Bianco, R.}, \bibinfo{author}{Errea, I.},
  \bibinfo{author}{Calandra, M.} \& \bibinfo{author}{Mauri, F.}
\newblock \bibinfo{title}{High-pressure phase diagram of hydrogen and deuterium
  sulfides from first principles: Structural and vibrational properties
  including quantum and anharmonic effects}.
\newblock \emph{\bibinfo{journal}{Phys. Rev. B}} \textbf{\bibinfo{volume}{97}},
  \bibinfo{pages}{214101} (\bibinfo{year}{2018})

\bibitem{Liu-strain}
\bibinfo{author}{Liu, C.} \emph{et~al.}
\newblock \bibinfo{title}{Strain-induced modulations of electronic structure
  and electron-phonon coupling in dense {H$_3$S}}.
\newblock \emph{\bibinfo{journal}{Phys. Chem. Chem. Phys.}}
  \textbf{\bibinfo{volume}{20}}, \bibinfo{pages}{5952--5957}
  (\bibinfo{year}{2018})

\bibitem{Zurek:2020b}
\bibinfo{author}{Cui, W.} \emph{et~al.}
\newblock \bibinfo{title}{Route to high-{T$_c$} superconductivity via {CH$_4$}
  intercalated {H$_3$S} hydride perovskites}.
\newblock \emph{\bibinfo{journal}{Phys. Rev. B}}
  \textbf{\bibinfo{volume}{101}}, \bibinfo{pages}{134504}
  (\bibinfo{year}{2020})

\bibitem{Sun:2020a}
\bibinfo{author}{Sun, Y.} \emph{et~al.}
\newblock \bibinfo{title}{Computational discovery of a dynamically stable cubic
  {SH$_3$}-like high-temperature superconductor at 100~{GPa} via {CH$_4$}
  intercalation}.
\newblock \emph{\bibinfo{journal}{Phys. Rev. B}}
  \textbf{\bibinfo{volume}{101}}, \bibinfo{pages}{174102}
  (\bibinfo{year}{2020})

\bibitem{Ge:2020a}
\bibinfo{author}{Ge, Y.}, \bibinfo{author}{Zhang, F.}, \bibinfo{author}{Dias,
  R.~P.}, \bibinfo{author}{Hemley, R.~J.} \& \bibinfo{author}{Yao, Y.}
\newblock \bibinfo{title}{Hole-doped room-temperature superconductivity in
  {H$_3$S$_{1-x}$Z$_x$ (Z=C, Si)}}.
\newblock \emph{\bibinfo{journal}{Mater. Today Phys.}}
  \textbf{\bibinfo{volume}{15}}, \bibinfo{pages}{100330} (\bibinfo{year}{2020})

\bibitem{Hu:2020}
\bibinfo{author}{Hu, S.}, \bibinfo{author}{Paul, R.},
  \bibinfo{author}{Karasiev, V.} \& \bibinfo{author}{Dias, R.}
\newblock \bibinfo{title}{Carbon-doped sulfur hydrides as room-temperature
  superconductors at 270 {GPa}}.
\newblock \emph{\bibinfo{journal}{arXiv preprint arXiv:2012.10259}}
  (\bibinfo{year}{2020})

\bibitem{quan2016van}
\bibinfo{author}{Quan, Y.} \& \bibinfo{author}{Pickett, W.~E.}
\newblock \bibinfo{title}{Van {Hove} singularities and spectral smearing in
  high-temperature superconducting {H$_3$S}}.
\newblock \emph{\bibinfo{journal}{Phys. Rev. B}}
  \textbf{\bibinfo{volume}{93}}~(10), \bibinfo{pages}{104526}
  (\bibinfo{year}{2016})

\bibitem{sano2016effect}
\bibinfo{author}{Sano, W.}, \bibinfo{author}{Koretsune, T.},
  \bibinfo{author}{Tadano, T.}, \bibinfo{author}{Akashi, R.} \&
  \bibinfo{author}{Arita, R.}
\newblock \bibinfo{title}{Effect of van {Hove} singularities on high-{T$_c$}
  superconductivity in {H$_3$S}}.
\newblock \emph{\bibinfo{journal}{Phys. Rev. B}}
  \textbf{\bibinfo{volume}{93}}~(9), \bibinfo{pages}{094525}
  (\bibinfo{year}{2016})

\bibitem{Ortenzi:2016}
\bibinfo{author}{Ortenzi, L.}, \bibinfo{author}{Cappelluti, E.} \&
  \bibinfo{author}{Pietronero, L.}
\newblock \bibinfo{title}{Band structure and electron-phonon coupling in
  {H$_3$S}: A tight-binding model}.
\newblock \emph{\bibinfo{journal}{Phys. Rev. B}} \textbf{\bibinfo{volume}{94}},
  \bibinfo{pages}{064507} (\bibinfo{year}{2016})

\bibitem{Akashi:2020}
\bibinfo{author}{Akashi, R.}
\newblock \bibinfo{title}{Archetypical ``push the band critical point''
  mechanism for peaking of the density of states in three dimensional crystals:
  Theory and case study of cubic {H$_3$S}}.
\newblock \emph{\bibinfo{journal}{Phys. Rev. B}}
  \textbf{\bibinfo{volume}{101}}, \bibinfo{pages}{075126}
  (\bibinfo{year}{2020})

\bibitem{Wang:2021}
\bibinfo{author}{Wang, T.} \emph{et~al.}
\newblock \bibinfo{title}{Absence of conventional room temperature
  superconductivity at high pressure in carbon doped {H$_3$S}}.
\newblock \emph{\bibinfo{journal}{arXiv preprint arXiv:2104.03710}}
  (\bibinfo{year}{2021})

\bibitem{Guan:2021}
\bibinfo{author}{Guan, H.} \& \bibinfo{author}{Liu, H.}
\newblock \bibinfo{title}{Superconductivity of light-elements doped {H$_3$S}}.
\newblock \emph{\bibinfo{journal}{arXiv preprint arXiv:2108.09437}}
  (\bibinfo{year}{2021})

\bibitem{Muller:2021}
\bibinfo{author}{M\"{u}ller, P.~C.}, \bibinfo{author}{Ertural, C.},
  \bibinfo{author}{Hempelmann, J.} \& \bibinfo{author}{Dronskowski, R.}
\newblock \bibinfo{title}{Crystal orbital bond index: Covalent bond orders in
  solids}.
\newblock \emph{\bibinfo{journal}{J. Phys. Chem. C}}
  \textbf{\bibinfo{volume}{125}}~(14), \bibinfo{pages}{7959--7970}
  (\bibinfo{year}{2021})

\bibitem{hoffmann1971theoretical}
\bibinfo{author}{Hoffmann, R.}
\newblock \bibinfo{title}{The theoretical design of novel stabilized systems}.
\newblock \emph{\bibinfo{journal}{Pure Appl. Chem.}}
  \textbf{\bibinfo{volume}{28}}~(2-3), \bibinfo{pages}{181--194}
  (\bibinfo{year}{1971})

\bibitem{hoffmann1970planar}
\bibinfo{author}{Hoffmann, R.}, \bibinfo{author}{Alder, R.~W.} \&
  \bibinfo{author}{Wilcox~Jr, C.~F.}
\newblock \bibinfo{title}{Planar tetracoordinate carbon}.
\newblock \emph{\bibinfo{journal}{J. Am. Chem. Soc.}}
  \textbf{\bibinfo{volume}{92}}~(16), \bibinfo{pages}{4992--4993}
  (\bibinfo{year}{1970})

\bibitem{Olah:1997a}
\bibinfo{author}{Olah, G.~A.} \& \bibinfo{author}{Rasul, G.}
\newblock \bibinfo{title}{From kekule's tetravalent methane to five-, six-, and
  seven-coordinated protonated methanes}.
\newblock \emph{\bibinfo{journal}{Acc. Chem. Res.}}
  \textbf{\bibinfo{volume}{30}}, \bibinfo{pages}{245--250}
  (\bibinfo{year}{1997})

\bibitem{Olah:1987}
\bibinfo{editor}{Olah, G.~A.}, \bibinfo{editor}{Prakash, G. K.~S.},
  \bibinfo{editor}{Williams, R.~E.}, \bibinfo{editor}{Field, L.~D.} \&
  \bibinfo{editor}{Wade, K.} (eds) \emph{\bibinfo{title}{Hypercarbon
  Chemistry}}  (\bibinfo{publisher}{John Wiley \& Sons}, \bibinfo{address}{New
  York}, \bibinfo{year}{1987})

\bibitem{Fahy:1987}
\bibinfo{author}{Fahy, S.} \& \bibinfo{author}{Louie, S.~G.}
\newblock \bibinfo{title}{High-pressure structural and electronic properties of
  carbon}.
\newblock \emph{\bibinfo{journal}{Phys. Rev. B}} \textbf{\bibinfo{volume}{36}},
  \bibinfo{pages}{36} (\bibinfo{year}{1987})

\bibitem{Pickard:carbon}
\bibinfo{author}{Martinez-Canales, M.} \& \bibinfo{author}{Pickard, C.~J.}
\newblock \bibinfo{title}{Thermodynamically stable phases of carbon at
  multiterapascall pressures}.
\newblock \emph{\bibinfo{journal}{Phys. Rev. Lett.}}
  \textbf{\bibinfo{volume}{108}}, \bibinfo{pages}{045704}
  (\bibinfo{year}{2012})

\bibitem{Daviau:2017a}
\bibinfo{author}{Daviau, K.} \& \bibinfo{author}{Lee, K. K.~M.}
\newblock \bibinfo{title}{Zinc-blende to rocksalt transition in sic in a
  laser-heated diamond-anvil cell}.
\newblock \emph{\bibinfo{journal}{Phys. Rev. B}} \textbf{\bibinfo{volume}{95}},
  \bibinfo{pages}{134108} (\bibinfo{year}{2017})

\bibitem{Hatch:2005a}
\bibinfo{author}{Hatch, D.~M.} \emph{et~al.}
\newblock \bibinfo{title}{Bilayer sliding mechanism for the zinc-blende to
  rocksalt transition in {SiC}}.
\newblock \emph{\bibinfo{journal}{Phys. Rev. B}} \textbf{\bibinfo{volume}{71}},
  \bibinfo{pages}{184109} (\bibinfo{year}{2005})

\bibitem{Gao:Si3C}
\bibinfo{author}{Gao, G.}, \bibinfo{author}{Liang, X.},
  \bibinfo{author}{Ashcroft, N.~W.} \& \bibinfo{author}{Hoffmann, R.}
\newblock \bibinfo{title}{Potential semiconducting and superconducting
  metastable {Si$_3$C} structures under pressure}.
\newblock \emph{\bibinfo{journal}{Chem. Mater.}} \textbf{\bibinfo{volume}{30}},
  \bibinfo{pages}{421} (\bibinfo{year}{2018})

\bibitem{Marx:1995a}
\bibinfo{author}{Marx, D.} \& \bibinfo{author}{Parrinello, M.}
\newblock \bibinfo{title}{Structural quantum effects and three-centre
  two-electron bonding in {CH$_5^+$}}.
\newblock \emph{\bibinfo{journal}{Nature}} \textbf{\bibinfo{volume}{375}},
  \bibinfo{pages}{216--218} (\bibinfo{year}{1995})

\bibitem{Lammertsma:1982}
\bibinfo{author}{Lammertsma, K.} \& \bibinfo{author}{Olah, G.~A.}
\newblock \bibinfo{title}{Diprotonated methane, {CH$_6^{2+}$}, and diprotonated
  ethane, {C$_2$H$_8^{2+}$}}.
\newblock \emph{\bibinfo{journal}{J. Am. Chem. Soc.}}
  \textbf{\bibinfo{volume}{104}}, \bibinfo{pages}{6851--6852}
  (\bibinfo{year}{1982})

\bibitem{Lammertsma:1983}
\bibinfo{author}{Lammertsma, K.} \emph{et~al.}
\newblock \bibinfo{title}{Structure and stability of diprotonated methane,
  {CH$_6^{2+}$}}.
\newblock \emph{\bibinfo{journal}{J. Am. Chem. Soc.}}
  \textbf{\bibinfo{volume}{105}}, \bibinfo{pages}{5258--5263}
  (\bibinfo{year}{1983})

\bibitem{Feng:2006a}
\bibinfo{author}{Feng, J.} \emph{et~al.}
\newblock \bibinfo{title}{Structures and potential superconductivity in
  {SiH$_4$} at high pressure: En route to ``metallic hydrogen''}.
\newblock \emph{\bibinfo{journal}{Phys. Rev. Lett.}}
  \textbf{\bibinfo{volume}{96}}, \bibinfo{pages}{017006 (1--4)}
  (\bibinfo{year}{2006})

\bibitem{Martinez2012}
\bibinfo{author}{Martinez-Canales, M.}, \bibinfo{author}{Pickard, C.~J.} \&
  \bibinfo{author}{Needs, R.~J.}
\newblock \bibinfo{title}{Thermodynamically stable phases of carbon at
  multiterapascal pressures}.
\newblock \emph{\bibinfo{journal}{Phys. Rev. Lett.}}
  \textbf{\bibinfo{volume}{108}}~(4), \bibinfo{pages}{045704}
  (\bibinfo{year}{2012})

\bibitem{LiuMa:2012b}
\bibinfo{author}{Liu, H.}, \bibinfo{author}{Zhu, L.}, \bibinfo{author}{Cui, W.}
  \& \bibinfo{author}{Ma, Y.}
\newblock \bibinfo{title}{Room-temperature structures of solid hydrogen at high
  pressures}.
\newblock \emph{\bibinfo{journal}{J. Chem. Phys.}}
  \textbf{\bibinfo{volume}{137}}, \bibinfo{pages}{074501 (1--7)}
  (\bibinfo{year}{2012})

\bibitem{materialsproject}
\bibinfo{author}{Sun, W.} \emph{et~al.}
\newblock \bibinfo{title}{The thermodynamic scale of inorganic crystalline
  metastability}.
\newblock \emph{\bibinfo{journal}{Sci. Adv.}} \textbf{\bibinfo{volume}{2}},
  \bibinfo{pages}{e1600225} (\bibinfo{year}{2016})

\bibitem{Allen:1975}
\bibinfo{author}{Allen, P.~B.} \& \bibinfo{author}{Dynes, R.~C.}
\newblock \bibinfo{title}{Transition temperature of strong-coupled
  superconductors reanalyzed}.
\newblock \emph{\bibinfo{journal}{Phys. Rev. B}}
  \textbf{\bibinfo{volume}{12}}~(3), \bibinfo{pages}{905}
  (\bibinfo{year}{1975})

\bibitem{perdew1996generalized}
\bibinfo{author}{Perdew, J.~P.}, \bibinfo{author}{Burke, K.} \&
  \bibinfo{author}{Ernzerhof, M.}
\newblock \bibinfo{title}{Generalized gradient approximation made simple}.
\newblock \emph{\bibinfo{journal}{Phys. Rev. Lett.}}
  \textbf{\bibinfo{volume}{77}}~(18), \bibinfo{pages}{3865}
  (\bibinfo{year}{1996})

\bibitem{kresse1996efficient}
\bibinfo{author}{Kresse, G.} \& \bibinfo{author}{Furthm{\"u}ller, J.}
\newblock \bibinfo{title}{Efficient iterative schemes for \textit{Ab Initio}
  total-energy calculations using a plane-wave basis set}.
\newblock \emph{\bibinfo{journal}{Phys. Rev. B}}
  \textbf{\bibinfo{volume}{54}}~(16), \bibinfo{pages}{11169}
  (\bibinfo{year}{1996})

\bibitem{blochl1994projector}
\bibinfo{author}{Bl{\"o}chl, P.~E.}
\newblock \bibinfo{title}{Projector augmented-wave method}.
\newblock \emph{\bibinfo{journal}{Phys. Rev. B}}
  \textbf{\bibinfo{volume}{50}}~(24), \bibinfo{pages}{17953}
  (\bibinfo{year}{1994})

\bibitem{Dronskowski:1993}
\bibinfo{author}{Dronskowski, R.} \& \bibinfo{author}{Bl{\"o}chl, P.~E.}
\newblock \bibinfo{title}{Crystal orbital hamilton populations ({COHP}):
  Energy-resolved visualization of chemical bonding in solids based on
  density-functional calculations}.
\newblock \emph{\bibinfo{journal}{J. Phys. Chem.}}
  \textbf{\bibinfo{volume}{97}}~(33), \bibinfo{pages}{8617--8624}
  (\bibinfo{year}{1993})

\bibitem{Maintz:2013}
\bibinfo{author}{Maintz, S.}, \bibinfo{author}{Deringer, V.~L.},
  \bibinfo{author}{Tchougr{\'e}eff, A.~L.} \& \bibinfo{author}{Dronskowski, R.}
\newblock \bibinfo{title}{Analytic projection from plane-wave and paw
  wavefunctions and application to chemical-bonding analysis in solids}.
\newblock \emph{\bibinfo{journal}{J. Comput. Chem.}}
  \textbf{\bibinfo{volume}{34}}~(29), \bibinfo{pages}{2557--2567}
  (\bibinfo{year}{2013})

\bibitem{togo2008first}
\bibinfo{author}{Togo, A.}, \bibinfo{author}{Oba, F.} \&
  \bibinfo{author}{Tanaka, I.}
\newblock \bibinfo{title}{First-principles calculations of the ferroelastic
  transition between rutile-type and {CaCl$_2$}-type {SiO$_2$} at high
  pressures}.
\newblock \emph{\bibinfo{journal}{Phys. Rev. B}}
  \textbf{\bibinfo{volume}{78}}~(13), \bibinfo{pages}{134106}
  (\bibinfo{year}{2008})

\bibitem{DFTraMO}
\bibinfo{author}{Yannello, V.~J.}, \bibinfo{author}{Lu, E.} \&
  \bibinfo{author}{Fredrickson, D.~C.}
\newblock \bibinfo{title}{At the limits of isolobal bonding: $\pi$-based
  covalent magnetism in {Mn$_{2}$Hg$_{5}$}}.
\newblock \emph{\bibinfo{journal}{Inorg. Chem.}} \textbf{\bibinfo{volume}{59}},
  \bibinfo{pages}{12304--12313} (\bibinfo{year}{2020})

\bibitem{giannozzi2009quantum}
\bibinfo{author}{Giannozzi, P.} \emph{et~al.}
\newblock \bibinfo{title}{Quantum espresso: A modular and open-source software
  project for quantum simulations of materials}.
\newblock \emph{\bibinfo{journal}{J. Phys. Condens. Matter.}}
  \textbf{\bibinfo{volume}{21}}~(39), \bibinfo{pages}{395502}
  (\bibinfo{year}{2009})

\bibitem{dal2014pseudopotentials}
\bibinfo{author}{Dal~Corso, A.}
\newblock \bibinfo{title}{Pseudopotentials periodic table: From {H} to {Pu}}.
\newblock \emph{\bibinfo{journal}{Comput. Mater. Sci.}}
  \textbf{\bibinfo{volume}{95}}, \bibinfo{pages}{337--350}
  (\bibinfo{year}{2014})

\bibitem{eliashberg:1960}
\bibinfo{author}{Eliashberg, G.~M.}
\newblock \bibinfo{title}{Interactions between electrons and lattice vibrations
  in a superconductor}.
\newblock \emph{\bibinfo{journal}{Sov. Phys. JETP}}
  \textbf{\bibinfo{volume}{11}}~(3), \bibinfo{pages}{696--702}
  (\bibinfo{year}{1960})

\end{thebibliography}

\section*{Data availability}
The data supporting this publication are available from the authors upon reasonable request.

\section*{Acknowledgments}
%
We acknowledge support from the U.S. National Science Foundation (DMR-1827815 to E.Z. and 
DMR-1933622 to R.J.H). This research was also supported by by the U.S. Department of Energy 
(DOE), Office of Science, Fusion Energy Sciences under Award No. DE-SC0020340 and DOE, 
National Nuclear Security Administration, through the Chicago/DOE Alliance Center under 
Cooperative Agreement Grant No.\ DE-NA0003975. Computations were carried out at the 
Center for Computational Research at the University at Buffalo (http://hdl.handle.net/10477/79221). 

\section*{Author contributions} 
R.J.H.\ and E.Z.\ conceived the project. X.W., T.B., K.P.H., and A.L.\ carried out the DFT calculations. All authors were involved in data analysis and results discussions. X.W., E.Z.\ and R.J.H.\ wrote the manuscript, with contributions from K.P.H.

\section*{Additional information}
\textbf{Supplementary Information} accompanies this paper at https://doi.org/xxx/aaa-bbb.ccc

\section*{Competing interests} 
The authors declare that no competing interests.

\end{document}